\documentclass[11pt]{article}

\usepackage{euscript}

\usepackage{latexsym}
\usepackage{amsmath, amssymb,graphics}
\usepackage[all]{xy}

\usepackage{fullpage}

\renewcommand{\(}{\begin{equation*}}
\renewcommand{\)}{\end{equation*}}
\newcommand{\bea}{\begin{eqnarray*}}
\newcommand{\eea}{\end{eqnarray*}}
\newcommand{\R}{{\mathbb R}}
\newcommand{\C}{{\mathbb C}}
\newcommand{\Z}{{\mathbb Z}}
\newcommand{\Q}{{\mathbb Q}}
\newcommand{\D}{{\mathbb D}}
\newcommand{\F}{{\mathbb F}}

\newcommand{\cA}{\ensuremath{\mathcal A}}
\newcommand{\cB}{\ensuremath{\mathcal B}}

\newcommand{\cH}{\ensuremath{\mathcal H}}

\newcommand{\cL}{\ensuremath{\mathcal L}}
\newcommand{\cM}{\ensuremath{\mathcal M}}
\newcommand{\cN}{\ensuremath{\mathcal N}}

\newcommand{\bo}{\raise-1mm\hbox{\Large$\Box$}}              % D'Alembertian
\def\H{\ensuremath{\ES{H}}}

\def\O{\ensuremath{{\cal O}}}

\newcommand{\cS}{{\mathcal S}}

\def\H{\ensuremath{\ES{H}}}

\def\D{\ensuremath{{\cal D}}}
\def\O{\ensuremath{{\cal O}}}

\newcommand{\beq}{\begin{equation}}
\newcommand{\eeq}{\end{equation}}

\numberwithin{equation}{section}

\renewcommand{\(}{\begin{equation}}
\renewcommand{\)}{\end{equation}}

\def\H{{\mathbb H}}

\def\R{{\mathbb R}}
\def\Z{{\mathbb Z}}
\def\Q{{\mathbb Q}}
\def\C{{\mathbb C}}

\def\1{{\bf 1}}

\def\<{\langle}
\def\>{\rangle}

\def\O{{\cal O}}

\numberwithin{equation}{section}

\renewcommand{\(}{\begin{equation}}
\renewcommand{\)}{\end{equation}}
\begin{document}

\begin{titlepage}
%\begin{flushright}

%hep-th/xxxxxxx
%\end{flushright}

\vspace{2em}
\def\thefootnote{\fnsymbol{footnote}}

\begin{center}
{\Large\bf 
On global anomalies in type IIB string theory}
\end{center}
\vspace{1em}

\begin{center}
Hisham Sati 
\footnote{e-mail: {\tt
hsati@pitt.edu}}
\end{center}

\begin{center}
Department of Mathematics\\
University of Pittsburgh\\
%139 University Place,\\
Pittsburgh, PA 15260
\end{center}

\vspace{0em}
\begin{abstract}
\noindent
We study global gravitational anomalies in type IIB string theory with 
nontrivial middle cohomology. This requires the study of the action of 
diffeomorphisms on  this group. 
Several results and constructions,
including some recent vanishing results via elliptic genera, 
make it possible to consider this problem. 
Along the way, we describe in detail the intersection pairing and the 
action of diffeomorphisms, and highlight  the appearance
of various structures, 
including the Rochlin invariant and its variants on the 
mapping torus.

\end{abstract}

\end{titlepage}

\tableofcontents

%%%%%%%%%%%%%%%%%%%%%%%%%%%%%%%
\section{Introduction}
%%%%%%%%%%%%%%%%%%%%%%%%%%%%%%%%%%%%%%%% 
There are two string theories with chiral supersymmetry in ten dimensions:
heterotic string theory and type IIB string theory. 
Due to the presence of chiral fermions, these theories might a priori 
suffer from anomalies, both local and global.  However,  
the first theory is in fact anomaly free, both locally \cite{AW} and 
globally \cite{W-global} \cite{W-Bar}. 
 The second theory is also free of 
local anomalies \cite{AW} as seen via a ``miraculous cancellation 
formula". Thus,  it then makes sense to discuss global anomalies.
The question of whether or not there are global anomalies in type IIB 
string theory has been investigated by Witten in Ref. \cite{W-global}, in the 
special case when the middle cohomology vanishes. 
These potential anomalies are gravitational since type IIB string theory
has no gauge fields, and hence there are obviously 
no global gauge anomalies. 
The aim of this paper is to investigate the question in general. 

\paragraph{Difficulties.}  
Witten's analysis indicates that type IIB string theory does not have global 
anomalies, but he states that his conclusion is not quite rigorous because of 
physical and mathematical uncertainties about how to treat 
antisymmetric tensor fields. This involves encoding the antisymmetric field
by an operator acting on bispinors. Furthermore, if the fifth Betti number of 
$X^{10}$ is not zero then this operator might have zero modes and hence 
might affect the result. Therefore, the paper \cite{W-global}
worked with the assumption that the fifth Betti number $b_5(X^{10})$ vanishes.  
In the case when $b_5(X^{10})\neq 0$, it was anticipated by Witten that 
the contribution of the self-dual tensor to a possible global anomaly
depends on the action of the diffeomorphism $f$ on the middle cohomology
group $H^5(X^{10};\R)$.

\paragraph{Recent developments.} There has been several developments since
the original treatment in \cite{W-global} which make it timely to revisit this problem. 
These developments include:

\vspace{2mm}
\noindent {\bf 1.} Much better understanding of the dynamics of self-dual fields
\cite{HT} \cite{SS} \cite{BHS} and their partition functions 
\cite{HNS} \cite{W-eff} \cite{Duality} \cite{HS} \cite{BeM} \cite{BeM2} \cite{Mon}.
The bosonic field appearing in the discussion of the anomaly is the 
5-form antisymmetric tensor, which is self-dual.

\vspace{2mm}
\noindent  {\bf 2.} More techniques for counting fermion zero modes 
\cite{At} \cite{DMW} \cite{KKT}.
 Global anomalies can involve the phase of the effective action, and 
can be investigated by counting the number of fermion zero modes
for the Dirac operator, and the number of zero modes for the signature 
operator. The latter has been studied in the dual context of M-theory \cite{S-sig}
\cite{S-corner}.
%While this will not appear explicitly in our discussion, it will be 
%a guiding conceptual point. 

\vspace{2mm}
\noindent {\bf 3.} An understanding of the need for a quadratic refinement of bilinear forms 
associated with the self-dual field \cite{W-eff} \cite{Duality}. 
Global anomaly considerations will have to take such refinements into account.
This will be central in our description of the structures associated with the anomaly. 

\vspace{2mm}
\noindent {\bf 4.} A better understanding of the geometry of diffeomorphisms 
\cite{ML} \cite{LMW} \cite{FK}. This includes the description of the holonomy
of the line bundle associated to the signature via the Rochlin invariant and 
variations thereof. 

\vspace{2mm}
\noindent {\bf 5.} A better understanding of structures related to 
the families index theorem using generalized cohomology \cite{Eb},
and corresponding vanishing theorems, using elliptic genera \cite{HL}.
The latter will be a major point in our description; it will allow us to deduce the
triviality of the holonomy of the anomaly line bundle. 

\vspace{3mm}
In addition to the above relatively recent works, we make use of classic 
results in topology not widely known in the physics literature; this 
includes \cite{Kr} \cite{BM}.

\paragraph{What we do.} 
%Most of what we do is conceptual. 
We provide the  context which
brings into light the relevance and the applicability of 
the above works, 
and apply the techniques  
in a suggestive way that leads us naturally 
 to arrive at the desired conclusions
on global anomalies of type IIB string theory in the 
case when $b_5(X^{10})\neq 0$.
In particular, we study the anomaly line bundles, their
holonomy, the effect of diffeomorphisms on the middle cohomology
in ten dimensions as well as on (almost) middle cohomology in 
eleven and twelve dimensions, via the mapping torus and its
bounding space. 
We view the main
point as a culmination of the above
works.
% which implies that
%\begin{theorem}
%\begin{center}
%{\it 
%Type IIB string theory is free of global anomalies.
%}
%\end{center}
%\end{theorem}
Along the way, we clarify the physical role of the various geometric, topological and 
algebraic structures involved. Thus the paper takes on 
an expository style throughout, and in certain sections is a survey.

\paragraph{Outline of the paper.}
We start in section \ref{sec rev} 
by reviewing the basic setting in type IIB string theory; this 
includes the field content, the self-dual fields and their local anomalies,
and the analysis of the global anomalies in the special case of 
vanishing middle cohomology. 
In section \ref{sec lb} we outline the construction of the line bundles
associated with the three relevant operators: the Dirac, Rarita-Schwinger,
and signature operators, using Atiyah's formulation of the latter. 
This then leads to a description of their 
holonomy in section \ref{sec par} in the context of Bismut and Freed. 
Having set up the holonomy in 
terms of eta invariants, we study the variation over the parameter space
in section \ref{sec glob}, thereby demonstrating cancellation via 
elliptic genera. Having spelled out the main ingredient in the 
anomaly cancellation, we go back and study details and tie some ends,
starting in section \ref{sec inter}, where we  include the middle 
cohomology 
and study the resulting intersection forms in ten and twelve
dimensions (with boundary) in section \ref{sec int10} and 
section \ref{sec int12}, respectively.  
The description of the action of diffeomorphisms on middle cohomology
is better done using the dual homology instead, which we explain
in section \ref{sec coh ho}. Then, in section \ref{sec quad}, we bring in 
quadratic forms and their refinements, which are essential for studying 
the self-dual field.
We distinguish quadratic forms appearing over $\Z$, $\Z_2$ and $\Q/\Z$ and 
in ten, eleven, and twelve dimensions in section \ref{sec quad z q}, where  
we also describe the connection to the Arf invariant. 
This leads to the study of characteristic vectors, in section \ref{sec char v},
 and Wu classes, in section \ref{sec Wu}, as
they also appear in the partition function, which we use for insight. 
Having set up algebraic, geometric and topological tools, we 
apply them to the study of diffeomorphisms in section \ref{sec dif}.
We consider diffeomorphisms preserving the Spin structure
and quadratic forms in section \ref{sec dif spin} and section 
\ref{sec diffq}, respectively. 
Finally in section \ref{sec Roch} we describe the relation to the
Rochlin invariant of the mapping torus, and in section \ref{sec Kreck}
to the Neumann, Fischer-Kreck, and Ochanine invariants.

\vspace{3mm}
Many of the constructions in this paper carry over to the M5-brane, for which
similar results hold. We plan to spell out the details elsewhere. 

\vspace{3mm}
{\it Note added.} After we finished writing this paper, a preprint appeared \cite{Mon2}
 in which the author announces a forthcoming work on the same problem.
  The two approaches seem to be different, 
 and we hope that they will 
 each enrich the knowledge in this area. A possible connection is that a
 good part of our (more formal) discussion can be
 recast in terms of theta functions and theta multipliers, via the topological
 interpretation of these in \cite{ML} \cite{LMW}.
 
%%%%%%%%
\section{Global anomaly cancellation}
%%%%%%%%%

In this section we start by reviewing the physical setting, then we set up the line
bundles  needed to study the global anomaly, and then we study the cancellation
of that anomaly.

%%%%%%%%%%%
\subsection{Review of the setting in type IIB string theory}
%%%%%%%%%
\label{sec rev}

We recall some of the basic aspects of type IIB string theory that we will
need for the rest of the paper and which will pave the way 
for the discussion of global anomalies.
We take type IIB string theory on a 10-dimensional Spin manifold 
$X^{10}$ with metric $g$, tangent bundle $TX$, and Spin bundle $S(X)$.

\vspace{3mm}
Type IIB supergravity
%\cite{GS}
 is the classical low energy limit of type IIB string theory. 
There is no manifestly Lorentz-invariant action for this theory 
\cite{Mar}, but one can write down the equations of motion 
\cite{Schw}\cite{HW}, and the symmetries and transformation rules 
\cite{Sch-W}.

\vspace{3mm}
A key property of a self-dual theory, like the type IIB theory, 
is that there is no single preferred 
action, but rather there is a family of actions parametrized by a Lagrangian
decomposition of the space of fields. In type IIB string theory there is 
no canonical choice of such Lagrangian decomposition for general spacetimes,
and that is why writing an action is difficult. However, in the case of 
product spacetimes and at low energy, corresponding actions can be written
\cite{BeM2}.

\paragraph{Field content.}
The field content of type IIB supergravity is:
%\begin{enumerate}

\vspace{2mm}
\noindent {\bf 1.}
{\it Bosonic:} metric $g$, two scalars $\phi$ and $\chi$,  
a complex 3-form field strength $G_3$ and a real self-dual 5-form
field strength $F_5$. Within this set, the latter field will be the main focus of this
paper. 

\vspace{2mm}
\noindent {\bf 2.} {\it Fermionic:} two gravitini $\psi^i$ ($i=1,2$) of the same chirality, i.e. 
sections of 
$S(X)^{\pm}\otimes(TX-2{\mathcal O})$ (with the same choice of sign), and 
two dilatini of the opposite
chirality, $\lambda^i \in \Gamma[S(X)^{\mp}]$. Here ${\cal O}$ 
denotes a trivial line bundle. 

%\end{enumerate}

\paragraph{Self-dual fields.} In $10=4\cdot 2+2$ dimensions, from a pair of spinors 
of the same chirality one can always construct the components of a 5-form $F_5$ by sandwiching 
five (different) $\gamma$-matrices between the two spinors (see e.g. \cite{AG}). 
There are two cases to consider, according to the signature of the 10-dimensional metric:

\vspace{2mm}
\noindent {\bf 1.} {\it  Lorentzian with metric $g^L$:} 
$F_5^L$ is self-dual if $F^L_{\mu_1\cdots \mu_5}=\frac{1}{5!} \epsilon_{\mu_1\cdots \mu_{10}}
F_L^{\mu_6\cdots \mu_{10}}$ with $\epsilon_{01\cdots 9}=+\sqrt{|g^L|}$ and is obtained from 
two spinors $\psi_I$ ($I=1,2$) satisfying 
$\gamma_M\psi_I=+\psi_I$, where $\gamma_M=\gamma_M^0 \cdots \gamma_M^9$ is the 
chirality matrix in Minkowski space. 

\vspace{2mm}
\noindent {\bf 2.}{\it Riemannian with metric $g^R$:} $F_5^R$ is called self-dual if 
$F^R_{j_1\cdots j_5}=\frac{i}{5!} \epsilon^R_{j_1\cdots j_{10}}F_R^{j_6\cdots j_{10}}$ with 
$\epsilon_R^{1\cdots 10}=1/{\sqrt{g^R}}$ and is obtained from two spinors
$\chi_I$ ($I=1,2$) satisfying 
$\gamma_E \chi_I=+\chi_I$, where $\gamma_E= i \gamma_E^1 \cdots \gamma_E^{10}$
is the chirality matrix in Euclidean space. 

\vspace{3mm}
\noindent With the careful conventions in Ref. \cite{BiM}, $\gamma_E=-\gamma_M$ upon 
analytic continuation, and so what is self-dual in one signature is 
anti-self-dual in the other. 
We will be mostly focusing on the Riemannian case for the geometric and topological 
considerations we have in mind.

%Hence one has the following table \cite{BiM}
%
% 
%\begin{center}
%\begin{tabular}{|c||c|c|}\hline
%        & Minkowski            & Euclidean    \\
%\hline
%\hline
%spinors & positive chirality &  negative chirality   \\
%\hline
%$5$-form& self-dual          & anti-self-dual    \\
%\hline
%\end{tabular}
%\end{center}

%\attn{below literal from \cite{BiM}}

\paragraph{Local anomalies with self-dual fields.}
In addition to arising from spinors, anomalies can result from a self-dual or
anti-self-dual $5$-form $F_5$ in ten dimensions. 
 Since $F_5$  can be constructed from a
pair of positive chirality spinors, the contribution to the 
anomaly is given by the $\hat{A}$-genus 
 multiplied by 
${\rm tr} \exp(iR)$, where
$R$ is the curvature of the metric $g$. 
There are two factors of $\frac{1}{2}$, one coming from chiral projection of the
spinor and another due to the fact that $F_5$ is real. 
 Overall, the index density is the degree twelve form
\begin{equation}
%{\rm ind}(iD_A)
I^A(R)
={1\over4} \left[\hat A(Z)
{\rm tr} \exp\left(iR\right)
\right]
={1\over4}[L(Z)]_{(12)}\;,
\label{indDA}
\end{equation}
where
$L(Z)$ is the Hirzebruch $L$-polynomial. 
The index of a negative chirality (anti-self-dual)
field is minus that of the corresponding positive chirality
(self-dual) field.
Therefore, the anomaly polynomial corresponding to (anti-)self-dual form field is
$
I^A=\left[ -\frac{1}{2}\frac{1}{4}L(Z)\right]_{(12)}$.
Then 10-dimensional type IIB supergravity with a self-dual 5-form field, a pair of chiral
spin $3\over 2$ Majorana-Weyl gravitinos, and a pair of anti-chiral Majorana-Weyl spin
$1\over 2$
fermions, leads to the total anomaly polynomial 
\(
I(R)=I^A(R) - I^{1\over 2}(R) + 
%\frac{1}{2}
I^{3\over 2}(R)\;.
\)
Here $I^{\frac{1}{2}}(R)$ is the $\hat{A}$-genus and 
$I^{\frac{3}{2}}(R)$ is the twisted $\hat{A}$-genus corresponding to the 
Rarita-Schwinger fields. The relative minus sign is due to the spinors being of
opposite chirality. 
%The factors of $1\over 2$ are included as we 
%have Majorana-Weyl spinors and not Dirac spinors.
Note that $I(R) = 0$ when all the terms are added, demonstrating that
 type IIB supergravity indeed has no local anomalies \cite{AW}.

\paragraph{Global gravitational anomalies for $b_5(X)=0$.}
Gravitational anomalies require working with the mapping torus $Y^{11}=(X^{10}\times S^1)_f$
of the 10-manifold $X^{10}$ corresponding to a diffeomorphism $f: X^{10} \to X^{10}$,
and then lifting to a bounding 12-manifold $Z^{12}$ with $Y^{11}=\partial Z^{12}$. 
Therefore, the study of anomalies in this case requires the use of the
index theorem for manifolds with
boundary, i.e. of Atiyah-Patodi-Singer (APS) type \cite{APS1} and hence involves
 eta invariants $\eta_D$, $\eta_R$, and $\eta_{\cS}$ 
 of the Dirac, the Rarita-Schwinger, and the signature operators, respectively. 
For a theory with $N_D$, $N_R$ and $N_S$ chiral Dirac, Rarita-Schwinger, and self-dual tensor 
fields, the change in the effective action under a diffeomorphism is \cite{W-global}
\begin{eqnarray}
\Delta I &=& \frac{\pi i}{2} \left( N_D \eta_D + N_R (\eta_R - \eta_D) -\frac{1}{2}N_S \eta_S \right)
\nonumber\\
&=& 2\pi i
\left(
\frac{1}{2}N_D {\rm index} (D) + \frac{1}{2}N_R ({\rm index}(R) - 2 {\rm index}(D)) 
-\frac{1}{8}N_S \sigma
\right)
\nonumber\\
&& - 
2\pi i \int_{Z} \left( 
\frac{1}{2}N_D \widehat{A} (R) + 
\frac{1}{2}N_R \left(K(R) - 2 \widehat{A}(R) \right)
-\frac{1}{8} N_S L(R)
\right)
\quad {\rm mod~} 2\pi i\;,
\end{eqnarray}
where $K(R)=I^{\frac{3}{2}}(R)$ 
is the Rarita-Schwinger index, $\sigma$ is the Hirzebruch signature,
and $\eta$ is the APS defect for each of the indicated operators
(cf. Section \ref{sec int12}). 
As indicated above, for type IIB string theory the values are 
$N_D=-1$, $N_R=2$, and $N_S=1$, so that 
\(
\Delta I = - 2\pi i \frac{\sigma(Z^{12})}{8} \quad {\rm mod~} 2\pi i\;.
\label{eq sig}
\)
The quantity $\Delta I$ is a topological invariant, since mod 16 
the signature $\sigma (Z^{12})$ depends only
on the topology of $\partial Z^{12}=(X^{10} \times S^1)_f$.
Therefore, if $\sigma(Z^{12})$ is divisible by 8 then the effective action is 
invariant and hence there are no global anomalies in this case \cite{W-global}.
As recalled in the introduction, the above analysis is done for the case when
 $b_5(X^{10})=0$. What we do in the rest of the paper is extend to the case when 
 there is nontrivial middle cohomology and then investigate the corresponding 
 effect of the relevant diffeomorphisms.

%%%%%%%%%%%%%%%%%%%%
\subsection{Line bundles on parameter space}
%%%%%%%%%%%%%%%%%%
\label{sec lb}

In this section we describe the line bundles on the parameter space which capture
the contribution to the global anomaly of each of our three operators. 
We will consider structures related to the situation depicted in this diagram
\(
\xymatrix{
X^{10}
\ar[r]
&
Y^{11}~
\ar[d]
 \ar@{^{(}->}[r]
 &
 Z^{12}
 \ar[d]
 &&
 \mathcal{N}
 \ar[d]
 &
 X^{10}
 \ar[l]
 \\
 &
 S^1~ 
  \ar@{^{(}->}[r]
&
\Sigma 
\ar[rr]
&&
\mathcal{B}
&
}
\label{diag 1}
\)
where

\noindent $\bullet$ $\Sigma$ is a Riemann surface with boundary $\partial \Sigma=S^1$.

\noindent $\bullet$ $Y^{11}=(X^{10}\times S^1)_f$ is the mapping torus corresponding 
to a diffeomorphism $f: X^{10} \to X^{10}$, which has the structure of a bundle 
 over $S^1$ with fiber the 10-manifold $X^{10}$. 

\noindent $\bullet$ $\mathcal{B}$ is the parameter space which will be the 
product of the intermediate Jacobian and the space of metrics modulo appropriate
diffeomorphisms. 

\noindent $\bullet$ $\mathcal{N}$ will be the total space of $\mathcal{B}$ with fiber
$X^{10}$. 

We will consider the above bundles with rescaled metrics, generically
$g_{\rm tot}=g_{\rm fiber} \oplus \frac{1}{\varepsilon^2} g_{\rm base}$ and 
take the limit $\varepsilon \to 0$.

\vspace{3mm}
A summary of 
the line bundles is provided by the following:

%\paragraph{Remarks on the line bundles.} We have the following 

\vspace{2mm}
\noindent {\bf 1.} {\it The line bundle corresponding to the Dirac and Rarita-Schwinger operators.}
These are line bundles over the space of metrics modulo diffeomorphisms. 
The index of the Dirac operator in 
 dimensions of the form $8k+4$, corresponding to anomalies in dimensions
$8k+2$, is even since the spinors are quaternionic. 
This implies that the first Chern class of the corresponding 
determinant line bundle will be even \cite{Fr0}.

\vspace{2mm}
\noindent {\bf 2.} {\it The line bundle corresponding to the signature operator.}
This is a line bundle over the space of the fields of middle degree, that is the
intermediate Jacobian Jac.  With $S^\pm$ the positive and negative 
chirality spinor bundles, 
the spaces $S^+ \otimes S^+$ and $S^-\otimes S^+$ are isomorphic to the 
spaces of even, respectively, odd self-dual forms. Therefore, 
a chiral Dirac operator coupled to the positive chirality Spin bundle 
$S^+$ can be viewed as a signature operator. 

\vspace{2mm}
\noindent {\bf 3.} {\it The combined line bundles.} 
We will consider the tensor product of the three line bundles above. 
However, two of them live on one space and the third lives in another.
One thought would be to take the product of the two spaces, the 
space of metrics modulo appropriate diffeomorphisms and the space of 
the fields of middle degree, and declare this as the general base space
of the line bundles. However, the restriction of the Hodge $*$ operator to 
middle degree cohomology provides a map from  
$\mathcal{M}/\mathcal{D}$ to  ${\rm Jac}$, which we can use to pull
back the line bundle over the latter. 

%with projections 
%${\rm pr}_1$ and ${\rm pr}_2$ to the first and second factors, 
%respectively. Then the total line bundle is the tensor 
%product $\cL^{\rm tot}={\rm pr}_1^*(\cL^{\rm Dir, RS})\otimes 
%{\rm pr}_2^*(\cL^{\rm sig})$ over the product 
%which -- for now-- we write schematically as 
%$\mathcal{M}/\mathcal{D}\times{\rm Jac}$.
%

\vspace{2mm}
We now consider the line bundles in more detail. Since the `usual' Dirac operator is 
familiar, we will focus on the signature operator (which itself can be viewed as
a Dirac operator), in the setting of diagram \eqref{diag 1}.

\paragraph{The line bundle corresponding to the signature operator.}
Let $\pi: X^{10} \to \cN \to \cB$ be a smooth fibration with fiber at a point
$x \in \cB$ a 10-manifold $X^{10}_x$ which is equipped with a metric
and a compatible Spin structure. The Spin structure varies smoothly over
the parameter space $\cB$ so that the structure group of the fibration
$\pi$ is a subgroup of the Spin diffeomorphism group. Then there is 
a principal Spin$(10)$ bundle $P(X^{10}_x)$ over the fibers. With 
$S^\pm$ the positive and negative chirality half-spinor representations
of Spin(10), we form the vector bundles $E_x^\pm = P(X^{10}_x) \otimes S^\pm$
and the corresponding Dirac operator
$D^A_x: L^2(E_x^\pm) \to L^2(E_x^\mp)$ on the Hilbert spaces of sections. 
As the parameter $x$ varies in $\cB$, the Hilbert spaces of sections $L^2(E_x^\pm)$ 
form Hilbert bundles $L^2(E^\pm)$ and the operators $D^A_x$ form a continuous
family of operators $D: L^2(E^\pm) \to L^2(E^\mp)$ on these Hilbert bundles. 

\vspace{3mm}
There is a well-defined complex line bundle $\det D^A$ over $\cB$. The fiber $(\det D)^A_x$
over a point $x \in \cB$ is isomorphic to the space 
$\left(\Lambda^{\rm max} \ker D^A_x \right)^* \otimes \left(\Lambda^{\rm max}~ {\rm coker} D^A_x \right)$.
There is a connection $\nabla$ on the line bundle 
$\det D^A$ over $\cB$
whose holonomy 
\footnote{We will describe the holonomy more fully in the next section.}
around an immersed 
circle $\gamma: S^1 \to \cB$ in the base manifold can be described as follows:
Pulling back by $\gamma$ there is an 11-dimensional manifold which is diffeomorphic to the
mapping torus $Y^{11}=(X^{10}\times S^1)_f$ with a diffeomorphism $f$ specified by $\gamma$.
Choosing an arbitrary metric $g_{S^1}$ on $S^1$, and using the projection
${\rm pr}: TY^{11} \to T_vY^{11}$ to the tangent bundle along the fibers, we obtain 
a Riemannian structure on $Y^{11}$. Since the structure group of the fibration $\pi$ is a subgroup of the 
Spin diffeomorphism group, it follows that $f$ is covered by a canonical Spin diffeomorphism 
\footnote{See Section \ref{sec dif spin} for details.} 
and the mapping torus has a natural Spin structure. 
From this Spin structure on $Y^{11}$ we obtain a Spin bundle over $Y^{11}$ with structure group
Spin(11) and a corresponding Dirac operator on the space of smooth sections of this bundle.

\paragraph{The signature of the extension of the mapping torus.}
Consider the fibration $X^{10} \to Z^{12} \to {\Sigma}$, where
$\Sigma$ is a Riemann surface with boundary. Assume that the total space
$Z^{12}$ is oriented; this is equivalent to assuming that 
the fundamental group $\pi_1(\Sigma)$ acts trivially on $H^{10}(X^{10})$. 
Then the signature $\sigma (Z^{12})$ on the middle cohomology
$H^6(Z^{12})$ is defined. 
Assuming appropriate metrics, we have an APS problem and the 
the signature is given by the APS index theorem.
The extension of the bundle structure 
from the mapping torus $Y^{11}$ to its bounding space $Z^{12}$ 
involves taking into account cobordism of diffeomorphisms, discussed
in section \ref{sec Kreck}.

\paragraph{The signature of the 10-manifold $X^{10}$.}
Consider the signature operator $\cS_X$ of the 10-manifold $X^{10}$ defined 
as $\cS_X=d+ d^\dagger$ : $\Omega^+(X^{10}) \to 
\Omega^{-}(X^{10})$, where $\Omega^\pm$ are the $\pm 1$-eigenspaces of the 
involution $\omega_p \to i^{p(p-1)+1} *\omega_p$ on a $p$-form $\omega_p$.
We see that for $p=5$, the involution is $\omega_5 \mapsto i *\omega_5$. 
Let $H^+$ and $H^-$ denote the solution spaces of 
$\cS_X u=0$ and $\cS_X^\dagger v=0$, respectively, i.e. the spaces
of harmonic forms in $\Omega^+(X^{10})$ and in $\Omega^-(X^{10})$.
Now if we vary $X^{10}$ over the fibers of $Z^{12} \to \Sigma$ we get a family
$D_x$ of Dirac operators and corresponding spaces 
$H_x^+$ and $H_x^-$ of harmonic forms which define vector bundles $H^+$ 
and $H^-$ over $\Sigma$. 
The Quillen line bundle $\cL$ is the bundle $\det H^-\otimes (\det H^+)^{-1}$ over
$\Sigma$ endowed with a natural unitary connection.

\paragraph{Zero modes of the signature operator.}
As explained in \cite{At}, one advantage of the signature operator 
over the generic Dirac operator is 
the ability of the former to control the integer ambiguity left by the
Bismut-Freed formulation.
This is because the zero eigenvalue of the Dirac operator cannot be
controlled in general, while for the signature operator the
identification of harmonic forms with cohomology via Hodge theory
fixes the integer ambiguity. 
The 0-eigenvalues of the signature operator, given by the harmonic 
bundles $H^\pm$, can be incorporated 
as follows (see \cite{At}).
Let $\cS'_X$ be the restriction of the signature oprator $\cS_X$ 
to the orthogonal complement of the harmonic spaces $H^\pm$. 
Then, via Quillen's formalism,
 $\det \cS'_X$ is a nowhere zero section of a line 
bundle $\cL'$ with a unitary connection over $\Sigma$. 
The harmonic bundles $H^\pm$ have natural metrics 
and connections induced via orthogonal projectiosn from the 
Hilbert space bundles of all forms. Then ${\cal H}= \det H^- \otimes (\det H^+)^{-1}$
is a line bundle with unitary connection. 
The two line bundles are then related as $\cL = \cL' \otimes \cH$ with the
induced unitary connection.

%\paragraph{How to extend the fibration 
%structure from $Y^{11}$ to $Z^{12}$.} 

%%%%%%%%%%%%%%%
\subsection{Holonomy of line bundles on the parameter space}
%%%%%%%%%%%%%%
\label{sec par}

All three operators that we have, namely the Dirac operator, the Rarita-Schwinger
operator, and the signature operator are of Dirac-type, that is are examples of
generalized Dirac operators. In this section we consider the holonomy  of the 
line bundles associated with these operators on the 
parameter space, using the general formulation of
 Bismut and Freed \cite{BF1} \cite{BF2}.

\paragraph{Holonomy of the line bundle.}
In order for the eta invariants to be independent of the metric on $S^1$, we rescale the 
metric on the circle $g_{S^1}$ as $\frac{1}{\epsilon^2}g_{S^1}$ and take the adiabatic limit,
given by  $\epsilon \to 0$. Corresponding to the rescaled metric we have a Dirac operator
$D^A_{\epsilon}$ on the mapping torus $Y^{11}$  and an eta invariant 
$\eta (D^A_\epsilon)$. We form the reduced eta invariant as 
$\overline{\eta}(D^A_\epsilon)=\frac{1}{2}(\eta (D^A_\epsilon) + \dim \ker D^A_\epsilon)
=\frac{1}{2}(\eta(D^A_\epsilon) + h(D_\epsilon^A))$, where $h$ is the number of zero modes.
Then the Bismut-Freed theorem  \cite{BF1} \cite{BF2}
says that the holonomy around the loop $\gamma$
of the connection $\nabla$ on the determinant line bundle is
\(
{\rm hol}(\gamma; \det D^A, \nabla)=\lim_{\epsilon \to 0} e^{-2\pi i \overline{\eta}(D^A_\epsilon)}\;.
\label{eq hol}
\)
The above has been for the signature operator (viewed as a generalized Dirac operator). 
There are similar results with obvious changes for the Dirac operator and the
Rarita-Schwinger operator; for the latter we have to replace the Spin bundle 
by the tangent bundle. Let us denote the resulting three lines bundles
with connections by $\cL_A$, $\cL_{RS}$, and $\cL_{\rm Dir}$, corresponding to the 
signature, the Rarita-Schwinger operator, and the Dirac operator, respectively. 
The holonomy of each of the connections on the line bundles corresponding 
to the three operators 
will have expressions of the form 
\eqref{eq hol}. The holonomy of the tensor product line bundle 
\footnote{Note that the signature is divisible by 8 (cf. section \ref{sec char v}),
which is `built into' $\cL_A$. 
See the remarks at the end of this section for more on this.
As cited at the end of the introduction, the new work 
\cite{Mon2} constructs line bundles explicitly from physical data. 
%However, the author points out that his line bundles do not lend themselves to 
%study of anomalies in a transparent way. We hope that the above
% bundles, although not constructed explicitly from first principles will be deemed
%useful for their transparent topological description.
}
\(
\cL_{\rm tot}:=\cL_A \otimes \cL_{RS}^{-1} \otimes \cL_{\rm Dir}^{4}
\)
with tensor product connection $\nabla^{\rm tot}$
will take the form 
\(
{\rm hol}(\gamma; \cL_{\rm tot}, \nabla^{\rm tot})=
\lim_{\epsilon \to 0} \exp \left\{-2\pi i \left[
\overline{\eta}( D^A_\epsilon) - 8\overline{\eta}( D^{RS}_\epsilon) 
+32\overline{\eta} (D_\epsilon)\right] \right\}\;.
\label{eq hol 3}
\)

\paragraph{Line bundles over $S^1$ vs. over $\Sigma$.}

\vspace{3mm}
The first Chern form of the Quillen line bundle
$\cL$ is \cite{BF1} \cite{BF2} 
\(
c_1(\cL)= -\frac{1}{2} \lim_{\varepsilon \to 0} 
\int_{X^{10}} L_{12}
\label{eq c1}
\)
 where the factor of $\frac{1}{2}$ arises because we are dealing with 
the $L$-polynomial rather than the $\hat{A}$-genus. 
When $\Sigma$ is the disk $\mathbb{D}^2$, the holonomy 
of $\cL$ around the bounding circle of $\Sigma$ is just 
$\exp(-\pi i \eta^0(Y^{11}))$, where $\eta^0=\lim_{\varepsilon \to 0} \eta^\varepsilon$
is the adiabtic limit of the eta invariant. 
For global anomalies we consider $\Sigma$'s that are
 topologically nontrivial. 
The extension from bundles over $S^1$ to bundles over $\Sigma$ will be discussed
in section  \ref{sec Kreck}.

\paragraph{Relative Chern class of the holonomy line bundle.}
As $\cL'$ (from the end of last section)
is trivialized by $\det \cS'_X$,
we have an isomorphism $\cL \cong \cH$ as a bundle but 
the isomorphism does {\it not} preserve the metric or connection. 
Using expression \eqref{eq c1}, 
the APS index formula can be written as 
\(
\sigma (Z^{12})=-2\int_\Sigma c_1(\cL) - \eta^0(Y^{11})\;.
\label{eq for sig}
\)
Since, via \cite{BF1},
$-\pi i \eta^0(Y^{11})$ is distinguished choice
for the logarithm of the holonomy of $\cL$ around $S^1=\partial \Sigma$,
we get a {\it relative} Chern class $c_1(\cL, \eta)$, where
as explained in \cite{At} the notation highlights that this Chern 
class is obtained from the eta invariant. Then  
\eqref{eq for sig} becomes 
\(
\sigma (Z^{12})= -2c_1(\cL, \eta)\;.
\)
This can be interpreted as signature of a local coefficient
system over $\Sigma$. The fibration $Z^{12}\to \Sigma$
gives a local coefficient system corresponding to the representation
of the fundamental group $\pi_1(\Sigma)$ on the cohomology of the 
fiber $H^*(X^{10})$. The middle cohomology $H^5(X^{10})$ 
gives a flat bundle with an antisymmetric form. 
This form can be changed to a Hermitian form by complexifying
coefficients and multiplying by $i$. This Hermitian form has 
type $(\frac{1}{2}b_5, \frac{1}{2}b_5)$, where 
$b_5=\dim H^5(X^{10})$ is the fifth Betti number of the fiber. 
From \cite{At}, multiplicativity of the signature for fiber bundles gives
that the signature of $\Sigma$ with coefficients in this 
flat bundle is equal to the signature of the total space 
\(
\sigma( \Sigma, H^5(X^{10}))= \sigma (Z^{12})\;.
\)
The contribution to $H^\pm$ from $H^j(X^{10})$ and 
$H^{5-j}(X^{10})$ for $j \neq 5$ cancel. That is, there 
is no contribution from the Ramond-Ramond fields other than 
the self-dual 5-form.

%\paragraph{The integer ambiguity.}
%\attn{analyze (4.11) in Atiyah}
%

\vspace{3mm}
As a warm-up for the general discussion in Section \ref{sec glob}
below, we illustrate some of the points on the variation of the above holonomy
in the case of change of Spin structure.

\paragraph{Different Spin structures.}
Suppose that our 10-manifold $X^{10}$ has more than one Spin structure
(see \cite{S2} for an extensive discussion of the effect of multiple Spin 
structures in the related context of M-theory). 
Suppose $\pi$ is a fibration of 10-manifolds $X^{10}$, with two preferred
Spin structures $\omega_1$ and $\omega_2$. Corresponding to these
two Spin structures there are families of Dirac operators
$D_{\omega_1}$ and $D_{\omega_2}$, and corresponding determinant 
line bundles $\det D_{\omega_1}$ and $\det D_{\omega_2}$. From the 
curvature formula of Bismut-Freed, these two complex line bundles 
have the same curvature 2-form. Hence, the contribution of their difference to
the local anomaly is zero, and hence the local anomaly is not sensitive to the
change of Spin structure. Stated more precisely, the complex line bundle 
$\cL_\omega= \det D_{\omega_1}/\det D_{\omega_2}= \det D_{\omega_1} \otimes (\det D_{\omega_2})^*$
is flat. However, for the global anomaly we need to investigate the holonomies of this
flat line bundle. Let $\omega'_1$ and $\omega'_2$ be the Spin structures on the 
mapping torus $Y^{11}=(X^{10}\times S^1)_f$ induced by the Spin structures 
$\omega_1$ and $\omega_2$ on $X^{10}$. The corresponding diffeomorphism 
$f$ is specified by the loop $\gamma$. Then, from \cite{LMW}, we have
that the holonomy of the the quotient line bundle $\cL_\omega$ is given by 
\(
{\rm hol}(\gamma; \cL_\omega, \nabla_\omega)=
\lim_{\epsilon \to 0} \exp \left\{-2\pi i \left[\overline{\eta}(D^\epsilon_{\omega_1'})
- \overline{\eta}(D^\epsilon_{\omega_2'})\right]
\right\}\;.
\label{eq hol omega}
\)
This formula holds for all three operators, namely the Dirac operator, 
the Rarita-Schwinger operator, and the signature operator. Hence,
the total variation with respect to the change of Spin structure
is given essentially by the product of three appropriate copies of 
 either side of expression \eqref{eq hol omega}. We need to consider the change
 of $\cL_{\rm tot}$ under the variation of Spin structure. This will then be
 (we use hol to denote ${\rm hol}(\gamma; \cL_{\rm tot},  \nabla^{\rm tot}_\omega)$)
 \bea
 \hspace{-5mm}{\rm hol}\hspace{-1cm}&&=
\lim_{\epsilon \to 0} 
\exp \left\{-2\pi i \left[
(\overline{\eta}( D^A_{\epsilon, \omega_1}) -  \overline{\eta}( D^A_{\epsilon, \omega_2})
-  8(\overline{\eta}( D^{RS}_{\epsilon, \omega_1}) -  \overline{\eta}( D^{RS}_{\epsilon, \omega_2}))
+ 32(\overline{\eta}( D_{\epsilon, \omega_1}) -  \overline{\eta}( D_{\epsilon, \omega_2})) \right]
\right\}
\nonumber\\
&=&
\lim_{\epsilon \to 0} 
\exp \left\{-2\pi i \left[
\left(\overline{\eta}( D^A_{\epsilon, \omega_1}) -  8\overline{\eta}( D^{RS}_{\epsilon, \omega_1}) 
+32 \overline{\eta}( D_{\epsilon, \omega_1})\right)
-
\left(\overline{\eta}( D^A_{\epsilon, \omega_2})
- 8  \overline{\eta}( D^{RS}_{\epsilon, \omega_2})
+    32\overline{\eta}( D_{\epsilon, \omega_2})\right)\right]
\right\}
\nonumber\\
%&=&
%\lim_{\epsilon \to 0} 
%\left(
%\exp \left\{-2\pi i [
%\left(\overline{\eta}( D^A_{\epsilon, \omega_1}) +  \overline{\eta}( D^{RS}_{\epsilon, \omega_1}) 
%- \overline{\eta}( D_{\epsilon, \omega_1})\right)]\}
%\exp \left\{-2\pi i [
%\left(\overline{\eta}( D^A_{\epsilon, \omega_2}) +  \overline{\eta}( D^{RS}_{\epsilon, \omega_2}) 
%- \overline{\eta}( D_{\epsilon, \omega_2})\right)]\}
%\right)
%\nonumber\\
&=& \lim_{\epsilon \to 0} 
\exp \left\{-2\pi i [ \overline{\eta}_{\rm tot}^{\omega_1} - \overline{\eta}_{\rm tot}^{\omega_2}]\right\}
\;,
\label{eq hol var}
 \eea
 where we have defined the `total reduced eta invariant'~ 
 $\overline{\eta}_{\rm tot}^{\omega_i}:= 
 \overline{\eta}( D^A_{\epsilon, \omega_i}) - 8 \overline{\eta}( D^{RS}_{\epsilon, \omega_i})
  +32\overline{\eta}( D_{\epsilon, \omega_i})$ for $i=1, 2$, corresponding to the two Spin structures.
 We will consider such a combination 
 again towards the end of next section. We will demonstrate, using the results of Ref. \cite{HL},
  that this combination of eta invariants
is constant under change of metric modulo appropriate diffeomorphisms.

%%%%%%%%%%
\subsection{The global anomaly cancellation}
%%%%%%%%%
\label{sec glob}

The local anomaly involves characteristic classes and characteristic forms.
The global anomaly will also involve curvatures of line bundles via the 
holonomy (see \cite{Fr} for an excellent general discussion). 
   Modular properties of characteristic forms following from elliptic genera 
   are powerful in giving relations among such forms. Along these lines, we will use the
   recent vanishing results of Ref. \cite{HL} throughout this section.
   That elliptic genera appear in a fundamental way  in type IIB string 
   theory is remarkable as it shows that they might have
   a role to play in type IIB, which is analogous to the role the Witten genus plays in 
 anomaly cancellation in heterotic string theory \cite{LNSW} and 
  in understanding topological aspects of M-theory \cite{S0} \cite{S1}.
 
\vspace{3mm}
Consider the mapping torus $Y^{11}=(X^{10}\times S^1)_f$ corresponding to a diffeomorphism 
$f: X^{10} \to X^{10}$ on the type IIB spacetime $X^{10}$, a Spin 10-manifold. 
Take this mapping torus 
to be the fiber in the smooth fiber bundle $Y^{11} \to \cM \to \cM_{\rm met}/\D$
over $\cM_{\rm met}/\D$, the quotient of the space of Riemannian metrics $\cM_{\rm met}$ 
 on $X^{10}$ by an appropriate diffeomorphism group $\D$. We will be interested in 
 $\D$ being the group of diffeomorphisms preserving the Spin structure on $X^{10}$
 and/or preserving the quadratic refinement corresponding to the self-dual 5-form field. 
  We will discuss such points extensively in section \ref{sec dif}.

    \vspace{3mm}
  Let $TY^{11}$ with metric $g_Y$ be the tangent bundle of the mapping torus viewed as the 
  vertical tangent bundle of the fiber bundle $\cM$. The total tangent bundle to $\cM$ splits
  orthogonally as $T\cM = T_H\cM \oplus TY^{11}$, where $T_H\cM$ is the smooth horizontal 
  subbundle. A metric $g_D$ on $T(\cM_{\rm met}/\D)$ can be lifted to a metric on $T\cM$ 
 which is the sum $g_D \oplus g_Y$. 
  
  \vspace{3mm}
  We need connections on the various spaces. First we start with the Levi-Civita connection
  $\nabla^L$ on the tangent bundle $T\cM$ of the total space, and then we form the 
  metric-preserving connection $\nabla^Y$ on the vertical tangent bundle $TY^{11}$
   defined by the relation  $\nabla^Y_UV=P_Y \nabla^L_UV$, for $U \subset  T\cM$ , 
   $V\subset TY^{11}$. Here $P_Y$ is the orthogonal projection from the total tangent 
   bundle $T\cM$ to the vertical tangent bundle $TY^{11}$. In order to consider characteristic 
   forms and characteristic classes we form the curvature $R^Y=(\nabla^Y)^2$ of the connection
   $\nabla^Y$. 
   
   \paragraph{The family signature operator.}
   Let $\{e_1, e_2, \cdots, e_{11}\}$ be an oriented orthogonal basis of $TY^{11}$.
  We can form the exterior bundle $\Lambda TY^{11}$ and consider differential forms
  on the mapping torus$Y^{11}$.
  Let $d^Y$ denote the exterior derivative along the fibers.
   Denote by $c$ the Clifford action on the complexified exterior algebra bundle 
   $\Lambda_\C(T^*Y^{11})$ of the cotangent bundle $T^*Y^{11}$ of the fiber. 
   On an element  $e$, this is given by $c(e)=e^*- i_e$, where $e^*$ is the 
   dual element in $T^*Y^{11}$ via $g_Y$ and $i_e$ is contraction with the 
   vector $e$. The chirality operator 
   $\Gamma=-c(e_1) \cdots c(e_{11})$ is a self-adjoint element satisfying
   $\Gamma^2={\rm Id}$. Define the family odd signature operator $\cS^Y$
   \(
   \cS^Y=\Gamma d^Y + d^Y \Gamma: C^\infty(\cM, \Lambda^{\rm ev}_\C(T^*Y^{11}))
   \to
   C^\infty(\cM, \Lambda^{\rm ev}_\C(T^*Y^{11}))\;.
    \) 
   For each point in the base space $x \in \cM_{\rm met}/\D$ corresponding to an 
   equivalence class of metrics, the restriction to the fiber over this point
   \(
    \cS^Y_x : C^\infty(Y^{11}_x, \Lambda^{\rm ev}_\C(T^*Y^{11})|_x)
   \to
  C^\infty(Y^{11}_x, \Lambda^{\rm ev}_\C(T^*Y^{11})|_x) 
   \)
   is the odd signature operator for the fiber $Y^{11}_x$ (cf. \cite{APS1}). 
   
   \paragraph{The family (twisted) Dirac operator.}
   Assume that $TY^{11}$ is Spin and form the Spin bundle $S(Y^{11})$.
   Consider the twisting of the Spin bundle by the complexified tangent bundle
   $V=T_\C Y^{11}$, that is    $S(Y^{11})\otimes T_\C Y^{11}$. On this twisted bundle 
   we have a connection $\nabla^V$ and a twisted Dirac operator 
   $D^Y \otimes V=\sum_{i=1}^{11} e_i \nabla^V_{e_i}$.  
   As in section \ref{sec par} above, for $x \in \cM_{\rm met}/\D$, let 
   $\eta_x(D^Y \otimes V)$ be the eta invariant
   corresponding to the twisted Dirac operator and consider the reduced 
   eta invariant 
   $\overline{\eta}_x(D^Y \otimes V)=\frac{1}{2}\left( 
   \eta_x(D^Y \otimes V) + \dim \ker(D^Y \otimes V)_x
   \right)$, as a function on $\cM_{\rm met}/\D$. 
   
   \paragraph{Consequences of modular invariance from elliptic genera.}
   We will review the results of \cite{HL} and provide an interpretation. 
   Let $\widetilde{T_\C Y^{11}}=T_\C Y^{11}-\dim T_\C Y^{11}$ be the reduced element 
   in K-theory of the total space $K(\cM)$  corresponding to the complexified
   vertical tangent bundle. Following \cite{HL}, define the $q$-expansion
   \(
   \Theta_2(T_\C Y^{11})=\bigotimes_{n=1}^\infty S_{q^n}(\widetilde{T_\C Y^{11}})\otimes 
   \bigotimes_{m=1}^\infty \Lambda_{-q^{m-\frac{1}{2}}}(\widetilde{T_\C Y^{11}}) \in 
   K(\cM)[[q^{\frac{1}{2}}]]\;. 
   \)
 The local anomaly cancellation formula can be written as 
 \(
 \{ L(TY^{1}, \nabla^Y)\}^{(12)}=8
 \sum_{r=0}^1 2^{6-6r}\left\{  
 \hat{A}(TY^{11}, \nabla^Y){\rm ch}(b_r(T_\C Y^{11}))
 \right\}^{(12)}\;,
 \label{eq loc}
 \)
  where  $b_r(T_\C Y^{11})$ are virtual vector bundles defined by the congruence
  \(
  \Theta_2(T_\C Y^{11})\equiv \sum_{r=0}^1 b_r(T_\C Y^{11})
  (8\delta_2)^{3-2r} \varepsilon_2^r \quad {\rm mod}~q\cdot K(\cM)[[q^{\frac{1}{2}}]]\;.
  \)
 Here $\delta_2$ and $\varepsilon_2$ are the modular forms
 written in terms of Jacobi theta functions with Fourier expansions
 in $q^{\frac{1}{2}}$ and are given by the expressions
 \bea
 \delta_2(\tau)&=& -\frac{1}{8} (\theta_1^4 + \theta_3^4)=- \frac{1}{8} - 3q^{\frac{1}{2}} - 3q + \cdots\;,
 \\
  \varepsilon_2(\tau)&=& \frac{1}{16} \theta_1^4 \theta_3^4=q^{\frac{1}{2}} +  8q + \cdots\;.
 \eea

\paragraph{Global anomaly cancellation via the family index.}
We have three family operators to consider: 
 The Dirac operator, the twisted Dirac operator, and the 
odd signature operator.
\footnote{As we mentioned earlier, the (family) signature operator itself can be viewed as a twisted
(family) Dirac operator.}   
Using \cite{Eb}, the family index of the odd signature operator
  on the oriented bundle  $Y^{11} \to \cM \to \cM_{\rm met}/\D$ is trivial, 
  that is, ${\rm ind}(\cS^Y)=0 \in K^1(\cM_{\rm met}/\D)$. Since 
  the integral over the fiber $\int_{Y^{11}}\hat{A}(TY^{11}, \nabla^Y) {\rm ch}(V, \nabla^V)$
  represents the odd Chern character of the index 
   ${\rm ind}(D^Y \otimes V)$, then the degree one class 
   $\left[ \int_{Y^{11}} L(TY^{11}, \nabla^Y)\right]$ is zero in de Rham cohomology. 
   The results of Bismut-Freed \cite{BF1} \cite{BF2} imply that
   \(
   d\{ \overline{\eta}_x(D^Y \otimes V)\}= \left\{ 
   \int_{Y^{11}}\hat{A}(TY^{11}, \nabla^Y) {\rm ch}(V, \nabla^V)
   \right\}^{(1)}\;.
   \label{eq d eta}
   \)
   Integrating both sides of \eqref{eq loc} over the fiber $Y^{11}$ gives 
   \(
   \left\{  \int_{Y^{11}}L(TY^{1}, \nabla^Y)\right\}^{(1)}-
   8\sum_{r=0}^1 2^{6-6r}
 \left\{  \int_{Y^{11}} \hat{A}(TY^{11}, \nabla^Y){\rm ch}(b_r(T_\C Y^{11}))
 \right\}^{(1)}=0\;,
   \)
   so that 
   \(
   d\{\overline{\eta}(\cS^Y)\}-8 \sum_{r=0}^1 2^{6-6r}d
   \{\overline{\eta}(D^Y \otimes b_r(T_\C Y^{11}))  \}
  =0\;. 
   \)
   Since $\cM_{\rm met}/\D$ is connected, this implies-- still applying \cite{HL}--
    that the combination 
  \bea
  \overline{\eta}_{\rm tot}&:=&\{\overline{\eta}(\cS^Y)\}-8 \sum_{r=0}^1 2^{6-6r}
   \{\overline{\eta}(D^Y \otimes b_r(T_\C Y^{11}))  \}
   \nonumber\\
   &=& \overline{\eta}(\cS^Y) - 8\overline{\eta}(D^Y \otimes T_\C Y^{11})
   +24 \overline{\eta}(D^Y)
   \label{eq eta tot}
   \eea
   is a constant function on the base $\cM_{\rm met}/\D$.
  Therefore, also the exponential 
  $\exp (2\pi i~ \overline{\eta}_{\rm tot})$
 is a constant function on the base. 
 That is, the phase is invariant under the variation of the metric modulo 
 (appropriately chosen) diffeomorphisms.  We interpret this as saying 
  that there are no global gravitational anomalies. 
 
 \paragraph{Remarks.} 
 {\bf 1.} In the above formal proof, there was nothing special about the base
 being explicitly $\cM_{\rm met}/\D$. In fact, the results hold for any 
 connected base. However, the choice we made is the one appropriate for 
 global anomalies in type IIB string theory. 
 
\noindent {\bf  2.} In addition, no detailed knowledge about the geometry of the base
 is needed. However, in order to illustrate the point, in the following sections
 we will include such
 aspects in order to describe the details of the anomaly cancellation
 in relation to the physical entities involved. 
 
\noindent {\bf 3.} We have left $\D$ generic for diffeomorphisms. We will be interested
 in diffeomorphisms which preserve the Spin structure and/or those 
 which preserve the quadratic refinements (the two diffeomorphisms
 are related). Again, in order to illustrate the process physically we will 
 describe such diffeomorphisms explicitly in section 
\ref{sec dif}.

\paragraph{The space of Riemannian metrics and its quotients.}
The space $\cM_{\rm met}$ of all Riemannian metrics $g_X$ on
$X^{10}$ is a contractible open cone inside the space 
$\Gamma (S^2T^*X^{10})$ of symmetric rank-2 tensor fields. 
The group ${\rm Diff}^+(X^{10})$ of orientation-preserving 
diffeomorphisms acts isometrically on $\cM_{\rm met}$. 
This action is free on the subset $\cM^{\rm noniso}_{\rm met}(X^{10})$
of metrics which admit no nontrivial isometries. 
See \cite{Ebi} for more details. If we insist on having a smooth
quotient then we should use this latter quotient for the moduli
space of metrics.

\paragraph{Remarks.}
{\bf 1.} The construction of the line bundle 
whose section is the partition function
is more involved since it is essentially 
Chern-Simons theory at level $\1\over2$ and hence 
requires taking delicate square roots
(see \cite{W-eff} \cite{Duality} \cite{HS} \cite{BeM} \cite{BeM2} \cite{Mon}).

\noindent {\bf 2.} For purposes of global anomalies one shows that
the phase of the form $e^{2\pi i \vartheta(x)/n}$ is constant over the 
moduli space of parameters, $x$. However, if $e^{2\pi i \vartheta(x)}$
is constant then so will be its $n$th roots for any $n$.
\footnote{In using such an argument, some torsion information will be lost. 
Since $c_1(\cL^n)=n c_1(\cL)$, it could happen that this is zero just because
$c_1(\cL)$ is an $m$-torsion class for $m$ a divisor of $n$. 
Therefore, our arguments work best when the Chern classes of 
the line bundles are not torsion. Such information requires working with K-theory,
which is beyond the scope of this paper.}

\vspace{3mm}
Having spelled out the main formal argument, we now turn to some of the details 
involving how the global anomaly cancels in our setting, as well 
as illuminating details involving the physics, and highlight
 some interesting consequences. 
 That is, even though the global anomaly cancellation did not care much about the details of the cancellation, it is nonetheless useful to see  how the anomaly cancels. 
 We view this as conceptually 
 analogous to the discussion in \cite{LNSW} in the case of 
 the heterotic string.  

%%%%%%%%%%%%
\section{Intersection pairings 
%and quadratic forms 
in 10, 11, and 12 dimensions}
%%%%%%%%%%%
\label{sec inter}

We will focus on the case $b_5(X^{10})\neq 0$, so that we have nontrivial 
cohomology $H^5(X^{10};\R)$. We will also consider extensions of this in two directions.
The first is to consider the lift to the mapping torus $Y^{11}=(X^{10}\times S^1)_f$ 
and to the bounding 12-manifold $Z^{12}$ and then study the corresponding cohomology
groups in these two other dimensions. The second  extension is to consider integral coefficients 
and separate the free and the torsion parts of the corresponding 
cohomology groups in all three relevant dimensions. We will see that our setting will
dictate preferences from the two sets of extensions.

%%%%%%%%%%
\paragraph{Identifying the intersection pairings in the relevant dimensions.}
%%%%%%%%%%%%%
Let $M$ be a closed oriented $m$-manifold. Define $T^k(M):=T^k(M;\Z)$ to be the 
torsion subgroup of the cohomology group $H^k(M;\Z)$, i.e.
\(
T^k(M)= H^k(M;\Z)_{\rm tors}=\{ \alpha \in H^k(M;\Z)~|~ rx=0~{\rm for~some~} r\in \Z \}\;.
\)
 The quotient 
 \footnote{In general there is a short exact sequence 
 $0 \to T^k(M) \to H^k((M; \Z) \to {\rm Fr}^k(M) \to 0$.}
${\rm Fr}^k(M)=H^k(M;\Z)/T^k(M)$ is then a free abelian group. 
The pairing 
\(
I:~ H^i(M;\Z) \otimes H^{m-i}(M;\Z) \to H^m(M;\Z)=\Z
\label{Gen}
\)
 induces a nonsingular 
pairing of free groups 
\(
I_F: ~{\rm Fr}^i(M) \otimes {\rm Fr}^{m-i}(M) \to \Z\;. 
\label{Free}
\)
There is also the nonsingular torsion pairing for $i\neq 0$
\(
L: T^i(M) \otimes T^{m+1-i}(M) \to \Q/\Z\;.
\label{Tors}
\)
Now we would like to concentrate on the cohomology of degrees 5 and 6
and, in the closed case, on spacetime dimensions 10 and 11; we would like to 
consider $X^{10}$ and its mapping torus $Y^{11}=(X^{10}\times S^1)_f$. 
In order to get an intersection form on middle cohomology of $X^{10}$, it 
is obvious that we have to look at  the pairing \eqref{Gen} or at
the pairing  \eqref{Free}.  This identifies for us 
the relevant pairings for $X^{10}$. 

\vspace{3mm}
Next, for $Y^{11}$, these two pairings do not give the correct degree,
but instead expression \eqref{Tors} does, due to the shift of one 
in degree. Therefore, 
in eleven dimensions we consider the torsion pairing
\(
L: T^6(Y^{11}) \otimes T^6(Y^{11}) \to \Q/\Z\;.
\label{eq L hom}
\)   
Of course we will also have the pairing on the free part, namely
${\rm Fr}^5(Y^{11}) \otimes {\rm Fr}^6(Y^{11}) \to \Z$.

\vspace{3mm}
Next we consider manifolds with boundary. Here our main case is the 
bounding 12-manifold $Z^{12}$ with $\partial Z^{12}=Y^{11}$, the mapping torus,
 and the main cohomology degree is 6. The cohomology pairing 
 $H^6(Z^{12};\Z) \otimes H^6(Z^{12}, Y^{11};\Z)
\to H^{12}(Z^{12}, Y^{11};\Z)=\Z$ defines a nonsingular pairing of free abelian 
groups ${\rm Fr}^6(Z^{12}) \otimes {\rm Fr}^6(Z^{12}, Y^{11}) \to \Z$. Since we are 
not particularly interested in degree 7 cohomology, we will not consider 
a torsion pairing for $Z^{12}$.

\vspace{3mm}
We would like to consider the symmetry of the relevant pairings identified above.
Useful references on bilinear and quadratic forms include \cite{MH} 
%\cite{Ra}
\cite{Sou}. 
we will need the following notions. 

\paragraph{Symmetry of quadratic forms over $\R$.}
For $\epsilon=+1$ or $-1$, an $\epsilon$-symmetric form $(V, \phi)$ 
is a finite-dimensional real vector space $V$ together with a bilinear pairing 
$\phi: V \times V \to \R$ sending $(x,y) \mapsto \phi(x,y)$ such that 
$\phi(x,y)=\epsilon \phi (y,x) \in \R$. 
The form is called {\it symmetric} for $\epsilon=+1$ and 
{\it symplectic} for $\epsilon=-1$. 
The pairing $\phi$ can be identified with the adjoint linear map to the 
dual vector space $\phi: V \to V^*={\rm Hom}(V,\R)$ sending 
$x$ to $(y\mapsto \phi(x,y))$ such that $\phi^*=\epsilon \phi$.
The form $(V, \phi)$ is nonsingular if $\phi: V \to V^*$ is an isomorphism. 
A {\it Lagrangian} of a nonsingular form $(V, \phi)$ is a subspace $L \subset V$ such that
 $L=L^\perp$, i.e.
 $L=\{x\in V~|~\phi(x,y)=0~{\rm for~all~} y {\rm~in~} L\}$. 
  The {\it hyperbolic $\epsilon$-symmetric form} is defined for any finite-dimensional 
 real vector space $L$ by
 $\mathbb{H}_\epsilon(L)=\left(L\oplus L^*, \phi=\binom{0~1}{\epsilon~0}\right)$,
 where $\phi: (L \oplus L^*) \times (L\oplus L^*) \to \R$ is given by 
 $((x,f), (y,g))\mapsto g(x) + \epsilon f(y)$ with Lagrangian $L$.  
 The inclusion $L \to V$ of a Lagrangian in a nonsingular $\epsilon$-symmetric 
form $(V, \phi)$ extends to an isomorphism 
$\mathbb{H}_\epsilon(L) \buildrel{\cong}\over{\longrightarrow} (V, \phi)$. 

\vspace{3mm}
The pairing on the middle cohomology of closed oriented $2k$-manifolds 
is symmetric for $k$ even and antisymmetric for $k$ odd. Therefore, 
in ten dimensions we will have symplectic forms corresponding to 
$\epsilon=-1$, and in twelve dimensions we will have symmetric forms
corresponding to $\epsilon=1$. The Lagrangian identifies the set of cohomology
classes for which the intersection form is zero. 

\vspace{3mm}
The cohomology of the pair $(Z^{12}, Y^{11})$ gives the following diagram, which
summarizes the relations
between the various cohomology groups we are considering
\(
\xymatrix{
%\ar[r]
%\ar[d]
&
T^6(Z^{12}, Y^{11})
\ar[r]^{\hspace{3mm} j}
\ar[d]
&
T^6(Z^{12})
\ar[r]^{ i}
\ar[d]
&
T^6(Y^{11})
\ar[r]^{\hspace{-2mm} \delta} 
\ar[d]
&
T^7(Z^{12}, Y^{11})
\ar[d]
\\
H^5(Y^{11}; \Z)
\ar[r]^{\hspace{-2mm} \delta^*}
\ar[d]
&
H^6(Z^{12}, Y^{11}; \Z)
\ar[r]^{\hspace{4mm} j^*}
\ar[d]
&
H^6(Z^{12}; \Z)
\ar[r]^{i^*}
\ar[d]
&
H^6(Y^{11}; \Z)
\ar[r]^{\delta^*}
\ar[d]
&
H^7(Z^{12}; \Z)
\ar[d]
\\
{\rm Fr}^5(Y^{11}) 
\ar[d]
\ar[r]^{\hspace{-3mm} \overline{\delta}}
&
{\rm Fr}^6(Z^{12}, Y^{11})
\ar[r]^{\hspace{2mm} \overline{j}}
\ar[d]
&
{\rm Fr}^6(Z^{12})
\ar[d]
\ar[r]^{\overline{i}}
&
{\rm Fr}^6(Y^{11})
\ar[r]^{\overline{\delta}}
\ar[d]
&
{\rm Fr}^7(Z^{12})
%\ar[d]
\\
0 
&
0
&
0
&
0
&
}.
\)
The maps $i$ and $\delta$ are adjoints of each other, and so are the 
maps $\overline{i}$ and $\overline{\delta}$, with respect to the pairings 
that we define in the following sections.
 We will study the cases in ten, eleven, and 
twelve dimensions in more detail.

%%%%%%%%%%%%
\subsection{Middle cohomology of closed 10-manifolds}
%%%%%%%%%%%%
\label{sec int10}

We now consider the degree five cohomology, as appropriate for the 5-form in type IIB string 
theory on $X^{10}$. 

\paragraph{The antisymmetric form over $\R$ of a closed 10-dimensional manifold.}
Consider a closed oriented 10-manifold $X^{10}$ with (co)homology 
with real coefficients. In this case the intersection form $\phi_X$ is  
defined using the fundamental class $[X^{10}]\in H_{10}(X^{10};\R)$ and is given by 
\(
 \phi_X: (x,y) \mapsto \langle x\cup y, [X^{10}]\rangle\;, \quad {\rm for~} x, y \in H^5(X^{10};\R)\;.
\)
The fact that $X^{10}$ is closed implies that the intersection form $\phi_X$ is 
nonsingular.

\paragraph{Classification of symplectic forms over $\R$.} It is natural to ask what possible
intersection pairings on $X^{10}$ can occur. These are characterized as follows
\begin{enumerate}
\item Every symplectic form $(V, \phi)$ over $\R$ is isomorphic to 
$\mathbb{H}_{-}(\R^p) \oplus \bigoplus_r (\R, 0)$ with $2p+r=\dim_\R V$. 
This form is nonsingular if and only if $r=0$. 
\item Two forms are isomorphic if and only if they have the same $p$ and $r$.
\item Every nonsingular symplectic form $(V, \phi)$ admits a Lagrangian
(as can be shown by induction on $\dim_\R V$). This implies that there are 
always cohomology classes whose pairing with every other class is zero. 
\end{enumerate}

\paragraph{Example 1.}
 The intersection form of $X^{10}=S^5 \times S^5$, the product of two 5-spheres,
  is the hyperbolic form 
 $\mathbb{H}_{-}(\R)=\binom{~0~~1~}{-1~0~}$. This example corresponds to 
$p=1$ and $r=0$ in the above classification. 

\vspace{3mm}
Let $Q$ be the intersection form over $\Z$ and let $b_5=\dim H^5(X^{10};\R)$. Then 
there exist $b_5 \times b_5 $ matrices $A$ and $B$ over $\Z$ for which 
$A^T B A=I_{b_5}$ is the identity matrix. Therefore, $\det Q=\pm 1$.
The free abelian group $H^5(X^{10};\Z)/T^5$, where $T^5$ is the torsion subgroup of the
integral cohomology group $H^5(X^{10};\Z)$, has a basis $\{ x_1, \cdots, x_{b_5}\}$ 
such that $x_i \cup x_j=\delta_{ij}\Lambda$, where
$\Lambda$ is a generator of the group $H^{10}(X^{10};\Z)$.

%%%%%%%%%%%%%%%%
\subsection{Intersection pairing on twelve-manifolds with boundary}
 %%%%%%%%%%%%%%%%
\label{sec int12}

The main operator we consider in the 12-dimensional case 
is the signature operator $\cS$. We now provide the setting and recall some of the 
basic properties that are relevant to our problem,
expanding on the remarks at the beginning of Section 
\ref{sec inter}. 
%We will concentrate on middle cohomology to 
%illustrate main point.  
For more background and details see Ref. \cite{Hir}
for the closed case and Ref. \cite{APS1} for the case with boundary.

\vspace{3mm} 
 Let $Z^{12}$ be a compact oriented twelve-manifold. Let $[\omega_1]$
and $[\omega_2]$ be elements of the middle cohomology group $H^6(Z^{12};\Z)$. 
The Hodge $*$-operator satisfies $*^2=1$ when acting on a 6-form in a twelve-manifold
$Z^{12}$, and hence $*$ has eigenvalues $\pm 1$.
The signature can be identified with the signature of the 
intersection pairing on $H^{6}(Z;\R)$; if we represent cohomology classes
$x$ and $y$ by closed forms 
$\alpha$ and $\beta$ then the intersection pairing is 
$\langle x, y \rangle=\int_Z \alpha \wedge \beta$.
Comparing with $\int_Z \alpha\wedge *\beta$, the $L^2$-inner product of 
$\alpha$ and $\beta$, we see that the intersection pairing is positive definite on the 
$+1$-eigenspace of $*$ and negative definite on the $-1$-eigenspace. 
%of $*$.  
Indeed, consider the bilinear form on middle cohomology
$
\phi_Z: H^6(Z^{12};\R) \times H^6(Z^{12};\R) \to \R
$,
defined by $\phi( [\omega_1], [\omega_2]):= \int_{Z^{12}} \omega_1 \wedge \omega_2$.
This has  the following properties

\vspace{2mm}
\noindent {\bf 1.} $\phi_Z$ is a $b_6 \times b_6$ symmetric matrix, where
$b_6=\dim H^6(Z^{12};\R)$. 

\vspace{2mm}
\noindent {\bf 2.}  $\phi_Z$ is nondegenerate since $\phi([\alpha], [\beta])=0$ for 
an $[\alpha]\in H^6(Z^{12};\R)$ implies $[\beta]=0$. 

\vspace{2mm}
\noindent {\bf 3.} 
 The definition of $\phi_Z$ is independent on the representatives
of $[\omega_1]$ and $[\omega_2]$. 

\vspace{2mm}
\noindent {\bf 4.} 
 Poincar\'e
duality implies that $\phi_Z$ has maximal rank. 

\vspace{2mm}
\noindent {\bf 5.} 
 On $Z^{12}$, $\phi_Z$ has real eigenvalues, $b_6^+$ of which are positive and 
$b_6^-$ of which are negative, with $b_6^+ + b_6^{-}=b_6$. The Hirzebruch signature is
defined as
$\sigma (Z^{12}):= b_6^+ - b_6^-$.

\vspace{3mm}
 Let
Harm${}^6(Z^{12})$ be the set of harmonic 6-forms on $Z^{12}$. Note that
Harm${}^6(Z^{12}) \cong H^6(Z^{12};\R)$ and each element of 
$H^6(Z^{12};\R)$ has a unique harmonic representative. There is 
a corresponding splitting of Harm${}^6(Z^{12})$ into $\pm1$-eigenspaces
$
{\rm Harm}^6(Z^{12}) = {\rm Harm}_+^6(Z^{12}) \oplus {\rm Harm}_{-}^6(Z^{12})$,
which block-diagonalizes $\sigma$; indeed for $\omega_6^\pm \in$
Harm${}_\pm^6(Z^{12})$, 
$
\phi_Z(\omega_6^+, \omega_6^+)=\int_{Z^{12}} \omega_6^+ \wedge \omega_6^+
=\int_{Z^{12}} \omega_6^+ \wedge * \omega_6^+=
(w_6^+, \omega_6^+) >0$,
where $(\omega_6^+, \omega_6^+)$ is the standard positive definite inner product on 
differential forms. Similarly, 
$
\phi_Z( \omega_6^-, \omega_6^-)= - \int_{Z^{12}} \omega_6^- \wedge * \omega_6^-
= -( \omega_6^-, \omega_6^-) <0$, and 
$\phi_Z( \omega_6^+, \omega_6^-)= - \int_{Z^{12}} \omega_6^+ \wedge * \omega_6^-
=-\int_{Z^{12}} w_6^- \wedge * w_6^+
= -( \omega_6^-, \omega_6^-) =0$.
Hence $\phi_Z$ is block-diagonal with respect to Harm${}_+^6(Z^{12}) \oplus$
Harm${}_-^6(Z^{12})$. Moreover, $b_6^\pm = {\rm dim}_\R {\rm Harm}_\pm^6(Z^{12})$.
Now $\sigma (Z^{12})$ is expressed as
\(
\sigma (Z^{12})= \dim {\rm Harm}_+^6(Z^{12}) -
\dim {\rm Harm}_{-}^6 (Z^{12})\;.
\)

\paragraph{Example 2: K\"ahler manifolds.}
If a compact K\"ahler manifold $Z$ is of even complex dimension, e.g. $6$, then the 
intersection pairing on the middle cohomology is of the form 
${\rm sign}(Z)=\sum_{p,q=0}^{6} (-1)^p h^{p,q}(Z)$, where $h^{p,q}(Z):=\dim H^{p,q}(Z)$ 
are the Hodge numbers. 

\paragraph{The signature index theorem.}
Poincar\'e duality shows that the Euler characteristic is given by 
$\chi (Z^{12})=b_6$ mod 2, so that $\sigma (Z^{12})=\chi(Z^{12})$ mod 2. 
Consider the operator ${\cal S}:=d + d^\dagger=d + *d*$. Since the (Hodge) Laplacian 
$\Delta={\cal S}^2$ is self-dual on $\Omega^*(Z^{12})$, the index of $\Delta$
vanishes identically. Also, ${\cal S}$ is self-adjoint, ${\cal S}={\cal S}^\dagger$, 
on forms $\Omega^*(Z^{12})$ and so ind(${\cal S})=0$ as well. 
However, a nontrivial complex is obtained when restricting 
${\cal S}$ to even forms;
$
{\cal S}^{\rm ev}: \Omega^{\rm ev}(Z^{12})^\C
\to 
\Omega^{\rm odd}(Z^{12})^\C$,
where $\Omega^{\rm ev}(Z^{12})^\C:=\bigoplus_i \Omega^{2i}(Z^{12})^\C$
and $\Omega^{\rm odd}(Z^{12})^\C:=\bigoplus_i \Omega^{2i+1}(Z^{12})^\C$.
The adjoint operator is 
$
{\cal S}^{\rm odd}:={\cal S}^{\rm ev \dagger}:
 \Omega^{\rm odd}(Z^{12})^\C
\to 
\Omega^{\rm ev}(Z^{12})^\C$.
Then the corresponding kernels are given by even and odd harmonic forms
$\ker({\cal S}^{\rm ev})=\oplus {\rm Harm}^{2i}(Z^{12})$, 
$\ker({\cal S}^{\rm odd})=\oplus {\rm Harm}^{2i+1}(Z^{12})$,
respectively.
As a result, the index calculates the Euler characteristic 
\(
{\rm ind}({\cal S}^{\rm ev})= \dim \ker({\cal S}^{\rm ev}) 
- \dim \ker({\cal S}^{\rm odd})=\chi (Z^{12})\;.
\)
For a complex-valued $r$-form
 $\omega \in \Omega^r(Z^{12})^\C$, application of the Hodge operator twice 
gives $**\omega=(-1)^r \omega$. Define a square root via the 
operator
$
\pi: \Omega^r(Z^{12})^\C \to \Omega^{12-r}(Z^{12})^\C
$
given by $\pi:=i^{r(r-1) + 6}*$ and which anticommutes with ${\cal S}$,
$\{\pi, {\cal S} \}= \pi {\cal S} + {\cal S} \pi=0$. 
Let $\pi$ act on $\Omega^*(Z^{12})^\C=\Omega^r(Z^{12})^\C$.
Since $\pi^2=1$, the eigenvalues of $\pi$ are $\pm1$. This gives a decomposition
of $\Omega^*(Z^{12})^\C$ into the $\pm1$-eigenspaces $\Omega^{\pm}(Z^{12})$ of
$\pi$ as $\Omega^*(Z^{12})^\C= \Omega^+(Z^{12}) \oplus 
\Omega^{-}(Z^{12})$. Since ${\cal S}$ anticommutes with $\pi$, the restriction of 
${\cal S}$ to $\Omega^+(Z^{12})$
defines the signature complex
$
{\cal S}_+: \Omega^+ (Z^{12}) \to \Omega^{-}(Z^{12})$,
where ${\cal S}_+:={\cal S}|_{\Omega^+(Z^{12})}$.
The index of the signature complex is 
\bea
{\rm ind} {\cal S}_+&=& \dim \ker({\cal S}_+) - \dim \ker({\cal S}_{-})
\nonumber\\
&=& \dim {\rm Harm} (Z^{12})^+ -
\dim {\rm Harm} (Z^{12})^-\;,
\eea
where ${\cal S}_{-}:={\cal S}^{+\dagger}:
\Omega^{-}(Z^{12}) \to \Omega^{+}(Z^{12})$ and 
Harm$(Z^{12})^\pm:=\{ \omega \in \Omega^\pm (Z^{12})~|~{\cal S}_\pm \omega=0 \}$.
Note that Harm${}^6(Z^{12})^\pm=$ Harm${}_\pm^6(Z^{12})$ since 
$\pi=*$ in Harm${}^6(Z^{12})$. 
The index theorem is \cite{Hir}
\(
{\rm ind} {\cal S}_+= \sigma (Z^{12})= \int_{Z^{12}} \left[L(TZ^{12})\right]_{(12)}\;.
\)

\paragraph{Classification of symmetric forms over $\R$.}
As in the antisymmetric 10-dimensional case, it is natural to ask which possible intersection forms
might arise in the 12-dimensional case. These can be characterized as follows: 
\begin{enumerate}
\item Every symmetric form $(V, \phi)$ is isomorphic to the direct sum
$\bigoplus_p (\R, 1) \oplus \bigoplus_q(\R, -1) \oplus \bigoplus_r(\R, 0)$
with $p+q + r=\dim_\R V$. The form $(V, \phi)$ is nonsingular if and only if $r=0$. 
\item Two forms are isomorphic if and only if they have the same nonnegaitve integers
$p, q$, and $r$.
\item The signature or index of $(V, \phi)$ is $\sigma(V, \phi)=p-q\in \Z$.

\item The following three conditions on a nonsingular forms $(V, \phi)$ are
equivalent 
\begin{enumerate}
\item $\sigma (V, \phi)=0$, that is $p=q$ (split signature). 
\item $(V, \phi)$ admits a Lagrangian $L$. 
\item $(V, \phi)$ is isomorphic to
 $\bigoplus_p (\R, 1) \oplus \bigoplus_p(\R, -1)\cong \mathbb{H}_+(\R^p)$. 
\end{enumerate}
\end{enumerate}

\paragraph{Example 3.}
The intersection form of the product of two 6-spheres, 
$Z^{12}=S^6 \times S^6$, is the symmetric hyperbolic form
$(H^6(S^6\times S^6;\R), \phi_Z)=\mathbb{H}_{+}(\mathbb{R})=\left( 
\R \oplus \R, \binom{0~1}{1~0}\right)$. 
Consequently, the signature is $\sigma(S^6\times S^6)=\sigma (\mathbb{H}(\R))=0$.
%The intersection form of $S^6 \times S^6$ is the 
%hyperbolic form $\mathbb{H}_{+}(\R)=\binom{0~1}{1~0}$.
This corresponds to the values $p=2$, $q=0$, and $r=0$, in the above
classification. 
%[Ranicki geometric and algebric]

\paragraph{The APS index for the case when $Z^{12}$ has nonempty boundary.}
Consider the signature operator ${\cal S}$ on the 12-manifold $Z^{12}$
when $\partial Z^{12}=Y^{11}$ is nonempty. In this case, in addition to the
Hirzebruch L-genus, one gets the corresponding eta invariant via 
the APS index theorem \cite{APS1}
\(
{\rm ind}({\cal S}_+)= \int_{Z^{12}} \left[L(TZ^{12})\right]_{(12)} - \eta_{\cal S}\;.
\) 
Unlike the characteristic form $L$, the invariant $\eta_{\cal S}$ is an 
analytic invariant. However, there are geometric ways of calculating 
this invariant without full knowledge of the spectrum of the 
operator (see \cite{S2} for a recent review in the related context
of M-theory). 

\paragraph{The symmetric form of a $12$-manifold with boundary over $\R$.}
Consider an oriented $12$-dimensional manifold $Z^{12}$ with
boundary $\partial Z=Y^{11}$ with 
(co)homology taken with real coefficients. 
The intersection form of $(Z^{12}, Y^{11})$ is the symmetric form 
given by the evaluation of the cup product on the fundamental class
$[Z^{12}]\in H_{12}(Z^{12}, Y^{11};\R)$, 
\(
\phi_Z: (x,y) \mapsto \langle x\cup y, [Z^{12}]\rangle  \quad {\rm for~} 
x, y \in  H^6(Z^{12}, Y^{11}; \R) \;.
\)
Poincar\'e duality and the universal coefficient theorem imply the relation 
between homology and cohomology in degree six
\(
H^6(Z^{12}, Y^{11}; \R)\cong H_6(Z^{12}; \R)\;, \quad H^6(Z^{12}, Y^{11}; \R)
\cong H_6(Z^{12}, Y^{11})^*\;.
\)
These groups fit into an exact sequence 
\(
\cdots 
\to
H_6(Y^{11}; \R) 
\to
H_6(Z^{12}; \R)
\buildrel{ \phi_Z}\over{\longrightarrow}
H_6(Z^{12}, Y^{11}; \R)
\to
H_5(Y^{11}; \R)
\to
\cdots .
\)
The isomorphism class of the intersection form is a homotopy invariant of 
$(Z^{12}, Y^{11})$. 

\paragraph{Example 4.} Take  $Z^{12}=\mathbb{D}^6 \times S^6$. This manifold
has a boundary $\partial{Z}=Y^{11}=S^5 \times S^6$, and then 
from the above relations we have
$H^6(\mathbb{D}^6\times S^6, S^5 \times S^6; \mathbb{R})
\cong H_6(\mathbb{D}^6 \times S^6; \R)$, as expected. 

\vspace{3mm}
Next we consider more fully the relation between homology and cohomology
in our context in the two other relevant dimensions, i.e. dimensions ten and eleven.

%%%%%%%%%
%\section{Ranicki}
%%%%%%%%%

%%%%%%%%%%%%%
\subsection{Cohomology vs. homology}
%%%%%%%%%%%
\label{sec coh ho}
For the purpose of connecting to geometry, and in particular for considering 
the action of the diffeomorphism group, it will be more convenient to work with 
homology rather than with cohomology. In this section we study how relevant 
properties of one group get translated to the other.  We make use of basic properties 
that can be found e.g. in \cite{Pr}.
The 12-dimensional case was considered towards the end of the previous section, so we
concentrate on the 10-dimensional case and also briefly on the 11-dimensional case.

\paragraph{(Co)homology as a module $V$ over a field.}
Let $M^{2n}$ be a closed oriented manifold and $\F$ an arbitrary field. 
Then $H_n(M^{2n}; \F)$ is an inner product space over $\F$, using the 
intersection number as inner product. The latter is  either symmetric 
(for $n$ even) or antisymmetric (for $n$ odd). Two cases 
are of particular interest to us:

\vspace{2mm}
\noindent {\bf 1.} {\it For $\F=\Z_2$, mod 2 coefficients}: If $x,y \in H_n(M^{2n};\Z_2)$ the intersection number is 
symmetric $\phi_M^*(x, y)=\phi_M^*(y, x) \in \Z_2$. Poincar\'e duality implies that 
the homology $H_n(M^{2n};\Z_2)$ is an inner product space over $\Z_2$.
 
 \vspace{2mm}
\noindent {\bf 2.} {\it For $\F=\Z$, integral coefficients}: 
The $\Z$-module ${\rm Fr}_n(M^{2n})=H_n(M^{2n};\Z)/\{\rm torsion~ subgroup\}$
 is an inner product space over $\Z$.

\vspace{3mm}
We start with ten dimensions. 

\paragraph{Dimension of homology.}
The matrix of the intersection form of the manifold $X^{10}$ 
is antisymmetric, and hence of even rank. Since the matrix is nondegenerate,
the rank should be equal to $\dim H_5 (X^{10};\R)$. Therefore, this dimension 
is even. In fact, this also follows from the fact that 
the Euler characteristic $\chi(X^{10})$ is even and that 
it has the same mod 2 value as the dimension of the middle cohomology.

%%%%%
%\attn{Prasolov Elements of homology theory}
%%%%%%%
\paragraph{Intersection pairings.}
Let $X^{10}$ be  a closed oriented 10-manifold. The 
intersection pairing on homology $\phi^*_X: H_i(X^{10}) \otimes H_{10-i}(X^{10}) \to \Z$
is given by $\phi^*_X(\alpha, \beta)=\langle PD^{-1}(\alpha), \beta \rangle$, where $PD^{-1}$ is the
inverse of the Poincar\'e duality isomorphism. Consider the restriction of 
$\phi^*_X$ to the free module ${\rm Fr}_5(X^{10})=H_5(X^{10};\Z)/{\rm Torsion}$. If we choose a basis 
for ${\rm Fr}_5(X^{10})$ then  
the intersection pairing is represented by an antisymmetric matrix whose
determinant is $\pm 1$, i.e. is a unimodular matrix. Such a pairing is 
called perfect.

\vspace{3mm}
We will concentrate on the case $i=5$ and work with more general coefficients. 
%Let $X^{10}$ be a closed oriented 10-manifold and 
Let $\F$ be any field. 
On the space  $H^5(X^{10};\F)$ we have seen that there is  
the bilinear form $\phi_X(x,y)=\langle x \cup y, [X^{10}]\rangle$, where 
$[X^{10}]$ is the fundamental class in $H_{10}(X^{10};\F)$. 
For the opposite orientation on $X^{10}$, i.e. taking $-X^{10}$, 
the fundamental class changes sign $[-X^{10}]=-[X^{10}]$.
Therefore, the bilinear form $\phi_X$  
changes sign as well: $\phi_{-X}=\phi_X$, where $\phi_{-X}$ is the bilinear form
for $-X^{10}$. 
On the dual space $H_5(X^{10};\F)$ there is the dual form 
$\phi_X^*(\alpha, \beta) = \langle \langle \alpha, \beta \rangle \rangle$, the intersection 
number. This also depends on the orientation as in the case for cohomology.

%\paragraph{$\R$ vs. $\Z$.}

%\vspace{3mm}
%Which to choose?

\paragraph{Relating integral homology and cohomology.}
 The universal coefficient theorem relates the homology groups 
$H_5(X^{10};\R) \cong H_5(X^{10};\Z) \otimes \R 
\cong \left( H_5(X^{10};\Z)/T_5\right) \otimes \R$. 
Let us consider this in more generality. 
Let $T_k(M)$ denote the torsion submodule of $H_k(M;\Z)$, i.e.
\(
T_k(M)= H_k(M;\Z)_{\rm tors}=\{ \alpha \in H_k(M;\Z)~|~ 
rx=0~{\rm for~some~} r \in \Z \}\;.
\)
Choose a complement ${\rm Fr}_k(M)$ of $T_k(M)$ in $H_k(M;\Z)$, i.e. a free 
submodule of $H_k(M;\Z)$ so that $H_k(M;\Z)\cong {\rm Fr}_k(M) \oplus T_k(M)$. 
Applying the universal coefficient theorem with $G=\Z$ gives the 
(noncanonical) isomorphisms
\(
H^k(M;\Z) \cong {\rm Fr}_k(M) \oplus T_{k-1}(M)\;.
\)
Note that the integral cohomology not only depends on on the free part of the 
homology in that degree but also, interestingly, 
on the torsion shifted down by one degree. 
We will make use of this in section \ref{sec dif}.

\vspace{3mm}
Now if we take $M$ to be an oriented $m$-manifold, then there is the 
Poincar\'e duality isomorphism $H_k(M;\Z) \cong H^{m-k}(M;\Z)$. 
Combining with the above symmetries gives 
the isomorphisms
\(
{\rm Fr}_k(M) \cong {\rm Fr}_{m-k}(M)\;, \quad T_k(M) \cong T_{m-k-1}(M)\;. 
\)
Hence, for $m=10$, $k=5$ we have the cohomology in terms of homology relation for $X^{10}$
\(
H^5(X^{10};\Z) \cong {\rm Fr}_5(X^{10}) \oplus T_4(X^{10})\;.
\label{uct 4}
\)
Therefore, we observe that if $H_4(X^{10};\Z)_{\rm tors}=0$ then middle integral 
cohomology is isomorphic  to the free part of the integral middle 
homology. If this happens  then there would be no torsion $(p,q)$ 
D3-branes. In the presence of such branes, however, one has to deal with
torsion 4-cycles. 

%\attn{Ok because we are restricting to the deg 5 field. Emphasize. 
%Then we are considering all fields towards the end.}

\paragraph{Example 5: Torsion in homology of degree four.} 
In light of equation \eqref{uct 4}, we need to get some idea about the torsion 
$T_4(X^{10})$ in $H_4(X^{10};\Z)$. 
Alternatively, we can look at $H_4(X^{10};\Z_p)$, that is degree-four homology with
coefficients in the cyclic group $\Z_p$ for $p$ a prime. If $H_4(X^{10};\Z)$ 
and $H_3(X^{10};\Z)$ are both finitely generated, e.g. if we
take them of the form 
\(
H_4(X^{10};\Z) \supset a(\Z) \oplus b(\Z_{p^k})\;, \quad 
H_3(X^{10};\Z) \supset c(\Z) \oplus d(\Z_{p^k})\;, ~~k\geq 1\;,
\)
then the universal coefficient theorem can be used (see \cite{Hat})
to show that 
\(
H_4(X^{10};\Z_p)\cong (a+b + d)\Z_p\;. 
\)
%Hatcher cor 3A.6

%\attn{Later we will see that: In \cite{LMW} $H_5(X^{10};\Z)$ 
%is assumed to be torsion-free. But this is Okay since our actual 
%starting point is cohomology, and so use UCT above}

\paragraph{Torsion and intersection forms.}
As we saw above, the bilinear forms $\phi_X$ and $\phi_X^*$ can be defined not only over a field 
$\F$ but also over the ring $\Z$. In this case, $H_n(X^{10})$ must be replaced by 
the free abelian group $H_5(X^{10};\Z)/T_5$, where $T_5$ is the torsion subgroup, 
because the intersection form vanishes on elements of finite order. 
The elements of finite order do not affect the intersection numbers: if 
$\alpha, \beta \in C_5(X^{10};\Z)$ and $r\alpha, s\beta \in C_5(X^{10};\R)$ 
are cycles, then the intersection forms are related as
$\langle \langle r\alpha, s\beta \rangle \rangle=rs \langle \langle \alpha, \beta \rangle \rangle$.
Therefore, the intersection forms over $\R$ and $\Z$ have the same matrix. This 
implies, in particular, that $H_5(X^{10};\R)$ has a basis in which the intersection 
form has integer coefficients. The significance of this for us is that torsion in ten 
dimensions will not need  to be considered. 
In fact, later when studying diffeomorphisms
we will assume $H_5(X^{10}; \Z)$ to be torsion-free, i.e. $T_5(X^{10})=0$.
However, we will see that is 
far from being the case in eleven dimensions. 
%\attn{why no torsion}

%%%%%%%%%%%%
\subsection{Quadratic forms and their refinements}
%%%%%%%%%%%%
\label{sec quad}

In this section we will consider quadratic refinements of the intersection 
forms that we encountered in the previous sections, mainly in Section
\ref{sec int10} and Section \ref{sec int12}. We start with the 
motivation and the go through a more detailed (and formal) description.

\paragraph{Quadratic functions from type IIB.}
The construction of the partition function for the self-dual 5-form field in
type IIB string theory requires the existence of a function 
$\Omega(x)$ from $H^5(X^{10};\Z)$ to the group $\Z_2=\{ \pm 1\}\subset U(1)$ obeying,
for all $x, y\in H^5(X^{10};\Z)$, the relation \cite{Duality} 
\(
\Omega (x+y)=\Omega (x) \Omega (y) (-1)^{x\cdot y}\;, 
\label{quad .}
\)
where $x\cdot y$ is the intersection pairing $\int_{X^{10}} x \cup y$. 
Furthermore, if we write $\Omega(x)=(-1)^{h(x)}$, then 
the mod 2 number $h(x)$ is given by 
$h(x)=\int_{Z^{12}}z \cup z$, where $z$ is a degree six cohomology class
in $H^6(Z^{12};\Z)$, extending $x$, 
with $Z^{12}$ the bounding Spin 12-manifold 
of the extension $Y^{11}$ of $X^{10}$ by a circle. 
When $Z^{12}$ is Spin, $h(x)$ is always even, so that there is no refinement, 
and hence no ambiguities in the partition function. 
However, Witten points out that it is more convenient to take $Z^{12}$ to be 
only oriented and not necessarily Spin. In this case, $h(x)$ is no longer 
necessarily well-defined mod 2, and the remedy for this
 is to replace the expression for
$h(x)$ by $\int_{Z^{12}} (z\cup z + v_6 \cup z)$, which is always even. 
Here $v_6$ is the 6th Wu class of $Z^{12}$ (see section 
\ref{sec char v} and section \ref{sec Wu}). If $z$ is taken to be a pull
back from $Y^{11}=\partial Z^{12}$ then $z\cup z$ vanishes for dimensional
reasons near the boundary, and the second summand also vanishes near the 
boundary because of the Spin condition $w_2=0$ there. This 
is put on firm mathematical ground by Hopkins and Singer \cite{HS}.

\paragraph{Remarks.} We state a few comments to help us proceed with the discussion.

\vspace{2mm}
\noindent {\bf 1.} The function $\Omega (x)$ above is written multiplicatively, i.e. 
using multiplication instead of addition. We note that when written 
additively, it coincides with the usual quadratic function, with the 
rule 
\(
q(x+y)=q(x) + q (y) + \phi_X(x,  y)
\)
replacing \eqref{quad .}, and with $\Omega$ replaced by $q$. 
 
\vspace{2mm}
\noindent {\bf 2.} The  above analysis for the partition function
requires a circle bundle, an instance of which is the product 
$Y^{11}=X^{10} \times S^1$. In comparison to our setting, 
this corresponds to the special case of the mapping torus
 with identity diffeomorphism.
 
 \vspace{2mm}
\noindent {\bf 3.}
 We will generally work with Wu-oriented manifolds.

\vspace{2mm}
\noindent {\bf 4.}
We will consider the relationship between classes on 
$Z^{12}$, classes on $Y^{11}$ and classes on $X^{10}$. This will 
be done both for `general' classes such as $z$ as well as `specific' 
classes such as the Wu class $v_6$.

\vspace{2mm}
\noindent {\bf 5.}
 In Ref. \cite{BeM} an approach was taken by looking at the bounding 
11-manifold to $X^{10}$ in order to study the partition  function 
of the self-dual theory. There, a choice of solution $\Omega$ is referred to 
as a choice of QRIF (Quadratic Refinement of the Intersection Form).
What we do here instead is a Chern-Simons construction in the sense of 
circle bundle then bounding.

\paragraph{Quadratic and bilinear forms.} 
Let $V$ be a finite-dimensional vector space over a field $\F$. A {\it quadratic form} on $V$ 
is a map $q: V \to \F$ satisfying

\vspace{2mm}
\noindent {\bf 1.} {\it Homogeneity in degree two}:
$q(ax)=a^2 q(x)$ for all $x$ in $V$ and $a$ in $\F$. 

\vspace{2mm}
\noindent {\bf 2.} {\it Polar identity}: The map $\varphi_q:V \times V \to \F$, defined by 
$\varphi_q(x,y)=q(x+y)-q(x)-q(y)$, is a bilinear form. This is called
the polar form of $q$. 
Note that if $\F$ has characteristic 2 then the polar form is automatically 
symmetric. 

\vspace{2mm}
\noindent The above relation in the second property 
can be `inverted' to give $q$ in terms of $\varphi$.
 We start with a bilinear form 
 $\varphi: V\times V \to \F$ is a bilinear form, and let $q_\varphi: V \to \F$
 be defined by $q_\varphi(x)=\varphi(x,x)$ for all $x$ in $V$. Then 
 $q_\varphi$ is a quadratic form with polar form $\varphi_{q_\varphi}=\varphi + \varphi^T$.

\paragraph{Working in a basis.}
Let $B=\{e_1, \cdots, e_n\}$ be an ordered basis for $V$. Then elements 
$x, y $ in $V$ have coordinate components $x=(x_1, \cdots, x_n), ~
y=(y_1, \cdots, y_n)$ in the basis $B$, and the bilinear form in this basis is
\(
\varphi (x,y)=\varphi(x_1e_1 + \cdots x_ne_n, y_1e_1 + \cdots + y_ne_n)=
\sum_{i,j}\varphi (e_i, e_j) x_i x_i\;.
\)
Then the matrix $[\varphi]_B:=(\varphi(e_i,e_j))$ on a given ordered basis 
completely determines the bilinear form. Consequently, in matrix notation,
we write $\varphi(x,y)=[x]^T [\varphi]_B [y]$. 
Two bilinear forms $\varphi$ and $\varphi'$ are isomorphic if and only if 
$[\varphi']_B= A [\varphi]_B A^T$ for some matrix $A \in GL(n,\Z)$.

\paragraph{Symmetric bilinear forms.}
If $\varphi$ is a symmetric bilinear form then the quadratic form associated to 
$\varphi$ is the function $q: V \to \Z$ defined by $q(x)=\varphi (x,x)$. 
A bilinear form over $\Z$ is called {\it even} (or {\it type II})
if $\varphi(x,x)$ is even for all 
$x$ in $V$. Since 
$\varphi(x+y, x+y)=\varphi(x,x) + \varphi(y,y) + 2\varphi(x,y)$, the bilinear 
form is even if and only if all elements on the diagonal, in the matrix description, 
are even. Note that, except in the case of characteristic 2, there is always 
an ordered basis for $V$ in which $\varphi$ is represented by a diagonal matrix.
A symmetric basis for $q$, or $\varphi_q$, is a basis $e_1, \cdots, e_n$ such that the associated matrix
$\varphi_q(e_i,e_j)$ has the generalized symmetric hyperbolic form 
$\mathbb{H}_{+}=\binom{0~~I}{I~~0}$.

\paragraph{Example 6.}
Let $a, b \in \F$. Consider the 2-dimensional quadratic form on $\F \times \F$ given by 
$q(x,y)=ax^2+xy+by^2$. The corresponding matrix for $q$ in the standard basis is
$A=\binom{a~~1}{0~~b}$, while the corresponding matrix for the polar form 
$\varphi_q$ is $\binom{2a~~1}{1~~2b}=A + A^T$.

\paragraph{Isotropic bilinear forms.}
A bilinear form $\varphi: V \times V \to \F$ is {\it isotropic} if $\varphi (x,x)=0$ for 
all $x$ in $V$. 
%\attn{This is confusing use of the word symplectic, do you have symm or antisymm?}
Note that 
$\varphi (x,y)+ \varphi(y,x)=\varphi(x+y,x+y)- \varphi(x,x) -\varphi(y,y)=0$, so that
every isotropic form is antisymmetric. The converse is not true in general. However, 
for $\F=\Z_2$, the converse holds since having $\varphi(x,x)=-\varphi(x,x)$ implies that
$\varphi(x,x)=0$. In fact, if $\F=\Z_2$ the bilinear form $\varphi$ is necessarily 
isotropic and it is always the case that $V$ possesses a symmetric basis. 
The first part of the fact can be seen from 
$\varphi_q(x,x)=q(2x) - 2q(x)=0$, since $2x=0\in V$ and $2q(x)=0\in \Z_2$.

%\vspace{3mm}
%Up to isomorphism, there is exactly one symplectic vector space in each even dimension.
%
%\attn{very simple classification as opposed to infinite in the other cases}
%
%\attn{Over $\Z_2$ it becomes symmetric automatically.}

%\subsection{generalities}

%%%%%%%%%%%%%
\subsection{Quadratic forms  on homology over $\Z_2$ and $\Q/\Z$ and the Arf invariant}
%%%%%%%%%%%%%%
\label{sec quad z q}
We have seen (cf. expression \eqref{quad .}) that the quadratic functions 
in type IIB string theory in ten dimensions take values in $\Z_2$. On the other 
hand, in eleven dimensions the relevant forms take values in $\Q/\Z$ 
(cf. expression \eqref{eq L hom}). 
In this section we provide further characterization of such forms. 

\vspace{3mm}
We consider a 10-dimensional Spin manifold $X^{10}$
and form the mapping torus $Y^{11}=(X^{10}\times S^1)_f$, which is 
an 11-dimensional Spin manifold. Then we form the 
bounding twelve-dimensional manifold $Z^{12}$.
We will consider the middle-dimensional homology of 
$X^{10}$ and study the corresponding `lifts' to $Y^{11}$ and 
to $Z^{12}$. We will also investigate what happens to the 
intersection pairing in the process. This is a homological analog of 
the discussion in Section \ref{sec int10} and Section \ref{sec int12}.

\subsubsection{Quadratic forms in ten dimensions}
\label{sec 10}

%\paragraph{Ten dimensions.}
Consider $X^{10}$, a closed Spin 10-manifold. 
Poincar\'e duality on homology with $\Z_2$ coefficients 
gives a nonsingular symmetric (since over $\Z_2$)
bilinear pairing 
\(
\phi_X^* : H_5(X^{10};\Z_2) \otimes H_5(X^{10};\Z_2) \to \Z_2\;.
\)
Using the construction in Ref. 
\cite{Br} we can define a quadratic refinement $q: H_5(X^{10};\Z_2) \to \Z_2$
of the $\Z_2$-intersection pairing $\phi^*_X$ which is essentially unique. 
Hence we can associate to each Spin manifold $(X^{10},\omega)$, with Spin 
structure $\omega$, the 
Arf invariant ${\rm Arf}(q)$ of $q$ called the {\it generalized Kervaire 
invariant}.

\paragraph{The Arf invariant.}
Let $\alpha_i, \beta_i$, for $i=1, \cdots, n$, be a symmetric basis for $q$, i.e.
\(
\varphi_q(\alpha_i, \beta_i)=\delta_{ij}, \quad 
\varphi_q(\alpha_i, \alpha_j)=\varphi_q(\beta_i, \beta_j)=0\;.
\)
Then the {\it Arf invariant} is defined as 
\(
A_q=\sum_{i=1}^n q(\alpha_i) q(\beta_i) \in \Z_2\;.
\)
If $B=\{ e_1, \cdots, e_n\}$ is a basis for the vector space $V$, then
any matrix $M$ such that $q(x)=x^TMx$ is called a matrix of $q$ with respect to $B$.
There is more than one possibility for the matrix $M$, but in the upper triangular 
form it is uniquely determined and given by the {\it normal form} 
$M=(m_{ij})$ with entries 
$m_{ij}$ equal to $q(e_i)$ for $i=j$, to $\varphi_q(e_i, e_j)$ for $i<j$ (above the 
diagonal), and to $0$ otherwise; that is we have the block-diagonal form  
$M=\binom{A~~I_n}{O_n~B}$, where $A={\rm diag}(q(\alpha_1), \cdots, q(\alpha_n))$,
$B= {\rm diag}(q(\beta_1), \cdots, q(\beta_n))$, $I_n$ is the $n\times n$ identity 
matrix, and $O_n$ is the $n\times n$ zero matrix. The Arf invariant can then be read off as
$A_q={\rm trace}(AB)$. For any other matrix   of 
$q$ in this same basis, say $\binom{A'~C}{D~~B'}$, we have 
${\rm trace}(AB)={\rm trace}(A'B')$ so that the Arf invariant can indeed be read 
from any matrix of $q$ in a symplectic basis.

\paragraph{Example 7: $V=\Z_2 \oplus \Z_2$.} 
%$b_5=1$ or 2?
 Let $\mathbb{H}_+$ be the hyperbolic space
with matrix $\binom{0~~1}{1~~0}$ on the basis $(\alpha, \beta)$. There are two quadratic forms
$q_i: \Z_2 \oplus \Z_2 \to \Z_2$, compatible with this bilinear form defined over 
$\Z_2$, given by 
\bea
\mathbb{H}_0: q_0(\alpha)=q_0(\beta)=0\;,
\\
\mathbb{H}_1: q_1(\alpha)=q_1(\beta)=1\;.
\eea
The two are manifestly not equivalent as quadratic forms. 
The vector space $\Z_2 \oplus \Z_2$ has only three nontrivial elements and is generated by
any two of them. Any change of basis $B=\{ \alpha, \beta \}$ to $B'=\{ \alpha', \beta' \}$ the relations
$\alpha'=\alpha$, $\beta'=\alpha' + \beta'$ hold after a possible change in the order of 
$\alpha$ and $\beta$. The new basis is still symplectic and the Arf invariant 
in this basis is $A_q=q(\alpha')q(\beta')$.  Using the transformation
and the fact that $q(\alpha + \beta)=\varphi(\alpha + \beta) + q(\alpha) + q(\beta)$ and 
$\varphi(\alpha, \beta)=1$, the Arf invariant takes the form
 $A_q=q(\alpha) + [q(\alpha)]^2 + q(\alpha)q(\beta)$. 
Now $q(\alpha) + [q(\alpha)]^2=2q(\alpha)=0$ in $\Z_2$ so that $A_q=q(\alpha)q(\beta)$, 
demonstrating that indeed the Arf invariant is well-defined for forms on $\Z_2 \oplus \Z_2$. 
Since $A_{\mathbb{H}_0}=0$ and $A_{\mathbb{H}_1}=1$ and these are the only quadratic forms
in two dimensions, this shows 
 that the Arf invariant completely classifies quadratic forms in dimension two. 
This is not the case when the dimension of the vector space, i.e. the 
rank of the middle cohomology, is greater than two.

%\attn{Following paragraph should go in the beginning with other UCT}

\paragraph{Consequence of $H_5(X^{10};\Z)$ being torsion-free.}
We will be interested in considering the case when
$H_5(X^{10};\Z)$ is torsion-free. Then by Poincar\'e duality the
homology group $H_4(X^{10};\Z)$ would  also be torsion-free. The 
universal coefficient theorem for homology
$H_n(M;G) \cong H_n(M;\Z) \otimes G \oplus {\rm Tor}(H_{n-1}(M;\Z),G)$
implies for $M=X^{10}$, $G=\Z_2$ and $n=5$, the isomorphism
$H_5(X^{10};\Z) \otimes \Z_2 \buildrel{\cong}\over{\longrightarrow} 
H_5(X^{10};\Z_2)$, under which the intersection pairing $\phi_X$ induces
a $\Z_2$-intersection pairing $\phi_X$ (mod 2). 
In particular, a quadratic refinement of this $\Z_2$-intersection pairing 
may be identified with a map $q: H_5(X^{10};\Z) \to \Z_2$ such that,
for all $x, y$ in $H_5(X^{10};\Z)$, 
\(
q(x+y)-q(x)-q(y)=\phi^*_X(x, y)~({\rm mod}~2)\;.
\label{eq q mod 2}
\)

%%%%%%%%%%%%%%%%%%%%%
\subsubsection{Quadratic forms in eleven dimensions}
%ntersection pairings in 10,11, and 12 dimensions}
%%%%%%%%%%%%%%%%%%%%%%%
\label{sec 11}

%\paragraph{Eleven dimensions.}
Consider the mapping torus $(Y^{11}, \omega')$, a compact Spin 11-manifold with Spin structure $\omega'$. 
Consider the torsion subgroup $T_5(Y^{11})$ of the homology group 
$H_5(Y^{11}; \Z)$.
Then we have a 
symmetric bilinear pairing, a homological counterpart of the cohomological
pairing \eqref{eq L hom},
\(
L:  T_5(Y^{11}) \otimes T_5(Y^{11}) \to \Q/\Z
\label{eq link}
\)
called the {\it linking pairing}, defined as follows. Given two classes
$y_1, y_2$ in $T_5(Y^{11})$, we represent them respectively by cycles
$\zeta_1$ and $\zeta_2$. Since these are torsion classes, there exists 
an integer $n$ such that $n \cdot \zeta_1$ is the boundary of a 6-chain $\xi$, 
that is $\partial \xi=n\cdot \zeta_1$. Define $L(y_1, y_2)$ by the formula
\(
L(y_1, y_2)=\left(\frac{1}{n}\right)\cdot ({\rm intersection~ number~of~}\xi {\rm ~and~}
\zeta_2)\;.
\) 
Poincar\'e duality and the universal coefficient theorem imply that 
this symmetric pairing $L$ is nonsingular. 
Corresponding to this pairing $L$ there is, via the general construction of 
Ref. \cite{BM}, the following quadratic refinement 
\(
Q_L: T_5(Y^{11}) \to \Q/\Z\;.
\label{eq QL}
\)
The {\it generalized Arf invariant} of $(Y^{11}, \omega')$ 
is defined by 
\(
{\rm Arf}(Y^{11}, \omega')=A(Q_L) 
\)
i.e. as the Arf invariant of the quadratic refinement $Q_L$. 
We will consider this invariant in the context of diffeomorphisms in
section \ref{sec dif}.

%\paragraph{Twelve dimensions.}

%%%%%%%%%%%%%%%
\subsection{Characteristic vectors and signature modulo 8}
%%%%%%%%%%
\label{sec char v}
In this section we provide an algebraic description of the signature 
modulo 8 appearing in equation \eqref{eq sig},  the formula
 for the global anomaly. The corresponding geometric 
aspects, together with the action of diffeomorphisms, will be discussed
in section \ref{sec dif}.

\vspace{3mm}
We will need the following definition. 
Let $V$ be a vector space over $\Z$. An element $v \in V$ is called {\it characteristic} if
$v\cdot x\equiv x\cdot x$ (mod 2) for every $x$ in $V$. 

\vspace{3mm}
In a basis, the definition of a characteristic is equivalent to the 
system of congruences 
$\sum_{j=1}^n a_{ij} v_j \equiv a_{ii}$ mod 2, for $i=1, \cdots, n$, where 
$(a_{ij})$ is the matrix representing the bilinear form $\varphi$ in the given basis. 
We can always find a characteristic by considering the stronger system of 
equations $\sum_{ij}a_{ij} v_j=a_{ii}$, for $i=1, \cdots, n$. This system will 
always have an integral solution since $\det(a_{ij})=\pm1$, and this 
solution is certainly a solution to the original congruence.

\paragraph{Existence of a characteristic element.}
From a vector space $V$ over $\Z$ we can form the induced 
vector space $V \otimes \Z_2$ over $\Z_2$. Let $\overline{x}$ denote the image
in $V \otimes \Z_2$ of the element $x$ in $V$ (that is, mod 2 reduction). 
Then the inner product $x\cdot y$ in $V$ gives rise to a 
$\Z_2$-valued inner product $\overline{x}\cdot \overline{y}=$\{residue class of 
$x\cdot y$ mod $2\}$ 
on $V\otimes \Z_2$. Since the function $V \otimes \Z_2 \to \Z_2$, given by 
$\overline{x}\mapsto \overline{x}\cdot\overline{x}$, is $\Z_2$-linear 
there is a unique element $\overline{v}\in V\otimes\Z_2$ 
which satisfies the equation $\overline{v}\cdot \overline{x}=\overline{x}\cdot \overline{x}$ for all
$\overline{x}$. Then the desired characteristic element is simply any 
preimage $v$ in $V$. 
Therefore, every vector space over $\Z$ possesses a characteristic element. 

\paragraph{Uniqueness of a characteristic element and the signature.} 
For any symmetric unimodular bilinear form on a lattice $\Lambda$ 
(i.e. a finite abelian group), let $\langle ~,~ \rangle$ be a symmetric unimodular pairing. 
Then all characteristic vectors in $\Lambda$ have the same square modulo 8 and they are
all equivalent modulo 2. Such a square is congruent to the signature modulo 8.
 If $v'$ is another characteristic element for $V$ then, by uniqueness of the 
mod 2 reduction, $v'$ is necessarily of the form $v'=v+2x$. Now the inner product
in $\Z$ of the new vector is 
$v'\cdot v'=v\cdot v + 4(v\cdot x + x\cdot v)$, which, upon using the definition of a characteristic 
element, is congruent to  $v\cdot v$ (mod 8). Hence the residue class of $v\cdot v$ modulo 8
is an invariant of the vector space $V$. This invariant is additive with respect to direct
sums. Now, for $p$ plus entries and $q$ minus entries, 
the signature is $\sigma=p-q$. Then if we form the orthogonal sum 
of $p$ copies of the inner product space $\langle 1\rangle$ and 
$q$ copies of the inner product space $\langle -1 \rangle$ and use the fact that 
$v\cdot v$ is congruent mod 8 to 1 and -1, respectively, on $\langle 1\rangle$ and 
$\langle -1 \rangle$, we get that 
$v \cdot v \equiv \sigma (V)$ (mod 8). This is called van der Blij's lemma,
and gives an algebraic understanding of expression \eqref{eq sig}.
For more arithmetic details, see Ref. \cite{Ser}.

\paragraph{Remarks.} 
{\bf 1.} {\it Constraints implied by the characteristic element.} When $v$ is 
a characteristic element then it can be checked that
 the determinant and the rank of the bilinear form
$\varphi$ are constrained by 
\(
 {\rm rank} \varphi + \det \varphi \equiv \varphi (v,v) +1 ~~\mod 4\;.
\label{eq const}
\)
In particular, if $\varphi$ is unimodular then the rank of $\varphi$ is given by the
value mod 4 of that bilinear form at the characteristic element. 
More general statements will be given in section \ref{sec diffq}.

\vspace{2mm}
\noindent {\bf 2.} {\it When the characteristic can be set to zero.}
For type II inner product spaces, we can safely set $v=0$. In this case, 
the signature is divisible by 8, so that $\frac{1}{8}\sigma(V)$ is an integer.

\paragraph{Insight from the partition function.} In \cite{W-eff} \cite{Duality} 
the partition function of the M5-brane was outlined. This was put on firm
mathematical ground in \cite{HS}. Constructing the partition function 
uses the fact that on an 8-manifold $M^8$ the expression
\(
\frac{1}{8}\int_{M^8} (\lambda^2-L(M^8))
\label{HS8}
\)
 is an integer, where $\lambda$ is the integral lift of the Wu class $v_4$.
  As we saw above, this has an algebraic explanation: the square of
the norm of a characteristic element of a non-degenerate symmetric bilinear form
over $\Z$ is always congruent to the signature mod 8.
For manifolds of dimension $4k$, the characteristic
elements for the intersection pairing in the middle dimension are the integer lifts
 $\lambda$ of the Wu class $v_{2k}$. The expression \eqref{HS8}
 is then an integer, and its variation
under to $\lambda \mapsto \lambda + 2x$ 
gives a quadratic refinement of the intersection pairing.
There are a lot of structural similarities between the M5-brane 
and type IIB string theory. 
The Chern-Simons construction for the partition function type IIB string theory 
amounts to  forming 
a circle bundle and then going to the bounding manifold and constructing 
the corresponding line bundle over the intermediate Jacobian. This 
construction for type IIB string theory requires the vanishing 
of the Spin cobordism group $\Omega_{11}^{\rm Spin}(K(\Z,6))$
of the Eilenberg-MacLane space $K(\Z, 6)$ representing the type IIB field in
degree five, conjectured to be the case in \cite{W-eff}. Witten's conjecture
is proved by Igor Kriz and the author in \cite{KS2}, thus 
allowing the applicability of the 
Hopkins-Singer construction to type IIB string theory.  
Indeed, 
the 12-dimensional version of expression 
 \eqref{HS8} was assumed in \cite{BeM}
to describe the Chern-Simons action in type IIB string theory.
This is also the basis of our discussion on the 
antisymmetric tensor field.
Note that self-duality was not an issue in arriving at the 
construction for the M5-brane \cite{W-eff} \cite{HS}, 
and hence we follow that line of thought
for type IIB string theory.

\vspace{3mm} For the Chern-Simons construction in type IIB, we need to 
consider the 6th Wu class $v_6$ on the 12-dimensional extension. One 
might wonder what will happen to  the 5th Wu class on $X^{10}$ itself.

\paragraph{The fifth Wu class.} Assuming that $H_4(X^{10})$ has no 2-torsion
(cf. Section \ref{sec coh ho}), then 
the Wu class $v_5$ vanishes if and only if there is a matrix representative for the 
intersection pairing so that all the diagonal entries are even. This happens if and 
only if every matrix representative for the intersection pairing 
has even diagonal entries. In fact, by the Wu formula, the odd degree class
$v_5$ is a composite class each of whose summands involves the first Stiefel-Whitney
class $w_1$; since we are dealing with oriented manifolds, 
$v_5$ will always be zero in the situations we consider.

%%%%%%%%%%%%%%%%%%%%%%
\subsection{Wu Structure via Spin structures}
%%%%%%%%%%%%%%%%%%%%%%
\label{sec Wu}

The study of Wu structures can be done in a very general setting
 with minimal topological structure and without the need for any
 geometry. Consider the topological space $B{\rm SO}[v_6]$ over BSO, the 
 classifying space for the stable orthogonal group, with 
fiber the Eilenberg-MacLane space $K(\Z_2, 5)$
\(
\xymatrix{
K(\Z_2, 5) 
\ar[rr]^=
\ar[d]
&&
K(\Z_2, 5)
\ar[d]
\\
B{\rm SO}[v_6] 
\ar[rr]
\ar[d]^\pi
&&
EK(\Z_2, 6)
\ar[d]
\\
B{\rm SO}
\ar[rr]^k
&&
K(\Z_2,6)
}\;.
\)
The $k$-invariant of this fibration is an element $v_6$ in the cohomology 
$H^6(B{\rm SO};\Z_2)$ defined by the 6th Wu class of the universal 
bundle $\xi$ over $B{\rm SO}$. 
A Wu structure on $X^{10}$ means a lifting $\tilde{\nu}: X^{10} \to B{\rm SO}[v_6]$ of the 
classifying map $\nu: X^{10} \to B{\rm SO}$ from $B{\rm SO}$ to the connected cover $B{\rm SO}[v_6]$,
that is there is a  diagram 
\(
\xymatrix{
&& 
B{\rm SO}[v_6]
\ar[d]^\pi
\\
X^{10}
\ar[rr]^\nu
 \ar@{-->}[urr]^{\tilde{\nu}}
&&
B{\rm SO}
}
\label{diag BSO}
\)  
such that $\pi \circ \tilde{\nu}=\nu$.

\vspace{3mm}
Let $\eta$ be a vector bundle  over our 10-manifold 
$X^{10}$ with  
vanishing first and second Stiefel-Whiteny classes $w_1(\eta)=w_2(\eta)=0$. Then 
by the Adem relations, the Wu class $v_6(\eta)$ is always zero. Therefore,
a Spin structure leads to a Wu structure. 
The situation is summarized in the following diagram 
\(
\xymatrix{
&&&& 
B{\rm SO}[v_6]
\ar[d]^\pi
\\
X^{10}
 \ar@{-->}[rrrru]
%\ar[rrrru]
\ar[rr]
\ar@/_1pc/[rrrr]_\nu
&&
BSpin
\ar[rr]
\ar[urr]_{\tilde{r}}
&&
B{\rm SO}
}
\)  
%\vspace{2mm}
\noindent The possible lifts $\tilde{r}$ are classified by $H^5(B{\rm Spin};\Z_2)$, which is zero. Therefore,
there is a unique lift and hence each Spin structure uniquely determines a Wu structure. 

\vspace{3mm}
Note that the Wu formula and Poincar\'e duality imply that the Wu 
class $v_6$ will vanish on all Spin 10-manifolds. 
Similarly this holds in eleven dimensions. 
\footnote{Later will consider relative Wu classes.} 
However, this is 
generally not the case in twelve dimensions.
Note that one might naively expect that $v_6$, being
 a middle cohomology class in twelve dimensions, will vanish in 
analogy to $v_5$ vanishing in ten dimensions (see end of 
Section \ref{sec char v}). However, this is not the case;
the main point is that  there is a big difference in the structure of Wu classes
in the even and odd degree cases.  
The appearance of $v_6$ in eleven and twelve 
dimensions will be discussed towards the end of 
section \ref{sec mid}; in fact there we will encounter a 
relative version of this class.

%%%%%%%%%%%%%
\section{Diffeomorphisms}
%%%%%%%%%%%%%
\label{sec dif}

In this section we consider diffeomorphisms  and their manifestation in ten, eleven, and 
twelve dimensions in detail, making use of the 
arguments and constructions in the previous sections.
Consider a diffeomorphim $f: X^{10} \to X^{10}$ which preserves some structure on 
the 10-manifold $X^{10}$. 
We certainly would like for $f$ to preserve the orientation on $X^{10}$. In addition,
we also would like to preserve further structure:

\vspace{2mm}
\noindent {\bf 1.} {\it The Spin structure}: We will consider Spin-preserving diffeomorphisms 
as well as the stronger notion of Spin-diffeomorphisms.
 
\noindent {\bf 2.} {\it  The quadratic refinement}: We would like for the diffeomorphisms to 
leave invariant the quadratic form coming from the middle cohomology (as described in 
previous sections).

\vspace{2mm}
\noindent Preserving the first structure is natural since $X^{10}$ is assumed to be a Spin manifold. 
The second structure is dictated by the fact that we are considering nontrivial 
middle cohomology involving such refinements. We will see that the above two types 
of diffeomorphisms are related, that is preserving a Spin structure is related to 
preserving the corresponding quadratic forms.
In summary, we would like to study the action of the diffeomorphism group on 

\vspace{2mm}
\noindent {\bf (i)} bilinear forms;

\noindent {\bf (ii)}  quadratic refinements;

\noindent {\bf (iii)}  middle cohomology.

\paragraph{The mapping torus of a diffeomorphism}
Let us temporarily abbreviate the 11-dimensional  
mapping torus $(X^{10}\times S^1)_f$ by $X_f$. 
If $f$ and $g$ are diffeomorphisms of $X^{10}$ then 
the cobordism class of the composition decomposes 
into classes  in 
$\Omega_{11}$, the cobordism group of closed oriented 
differentiable 11-manifolds,
as $[X_{f\cdot g}]=[X_f]+[X_g]$.

%%%%%%%%%
\subsection{Diffeomorphisms preserving the Spin structure}
%%%%%%%
\label{sec dif spin}

%\paragraph{Spin diffeomorphisms}.
Consider a 10-manifold $X^{10}$ with frame bundle $F(X)$ and Spin bundle 
$S(X)$ with structure groups SO(10) and Spin(10), respectively. 
Given an orientation-preserving  diffeomorphism $f: X^{10} \to X^{10}$,
the differential $df$ of $f$ gives a diffeomorphism 
at the level of the frame bundle $df: F(X) \to F(X)$, and hence
an isomorphism $(df)^*:H^1(F(X);\Z_2) \to H^1(F(X);\Z_2)$. 
Such a diffeomorphism 
$f$ preserves the Spin structure $\omega$ if $(df)^*(\omega)=\omega$ in 
$H^1(F(X);\Z_2)$. This is also called a {\it Spin preserving diffeomorphism}.
On the other hand, a Spin diffeomorphism $\hat{f}$ of $(X^{10}, \omega)$ is a pair
$\hat{f}=(f,b)$ consisting not only of a Spin preserving diffeomorphism $f$ but also
of a bundle map $b: S(X) \to S(X)$ covering $f$; then there is  a   
%This is equivalent to the existence of a bundle map 
%$b: S(X) \to S(X)$ making the following diagram 
commutative diagram
\(
\xymatrix{
S(X) 
\ar[rr]^b
\ar[d]
&&
S(X)
\ar[d]
\\
F(X) 
\ar[rr]^{df}
&&
F(X)
}\;.
\)

\paragraph{Spin diffeomorphisms.}
A {\it Spin diffeomorphism}  is a quadruple $(X^{10}, \omega, f, h)$ where \cite{Kr}
\cite{LMW} \cite{FK}

\noindent {$\bullet$} $w: X^{10} \to B{\rm Spin}$ is a Spin structure. 

\noindent {$\bullet$} $f: X^{10} \to X^{10}$ is a diffeomorphism.

\noindent {$\bullet$} $h: I \times X^{10} \to B{\rm Spin}$ is a Spin structure on $[0, 1] \times X^{10}$ 
such that $h_0=\omega$ and $h_1=\omega\circ f$. 

\vspace{3mm}
\noindent For a given diffeomorphism $f$ with this property there are exactly 
two homotopy classes of choices for $h$ since 
$H^1(I \times X^{10}, \partial I \times X^{10};\Z)\cong \Z_2$. 
%Denote by $\omega_h$ the Spin structure given by $h$. 
Because of the double 
covering map Spin $\to$ SO, the following map is also two-to-one 
\(
\{ {\rm Spin~diffeomorphisms}\} 
\buildrel{2:1}\over{\longrightarrow} \{{\rm Diffeomorphisms~ preserving ~Spin ~structures}\}\;.
\)
Therefore, as far as Spin structures are concerned, we can have two 
quotients of the space of metrics ${\cal M}_{\rm met}$
on $X^{10}$, namely 
$$
{\cal M}_{\rm met}/\{ {\rm Spin~diffeomorphisms}\}~~{\rm  and}~~ 
{\cal M}_{\rm met}/ \{{\rm Diffeomorphisms~ preserving ~Spin ~structures}\}\;.
$$

\paragraph{The mapping torus of a Spin diffeomorphism.}
One way of defining the mapping torus 
$Y^{11}=(X^{10}\times S^1)_f$
in this case is to take 
(cf. \cite{FK}) $(Y^{11}, \omega)$ to be the Spin manifold formed as follows:
$\Z$ acts on $\R\times X^{10}$ by 
$(n, (r,x))\mapsto (r-n, f^n(x))$ and then $Y^{11}:=\R \times_\Z X^{10}$, with the 
Spin structure $\omega_h$ induced by $h$.

 \paragraph{Example 8.}
Consider $X^{10}=S^5 \times S^5$ with the Spin structure given by the stable 
trivialization of the tangent bundle 
$
TX^{10} \oplus \O^2=(TS^5 \oplus \O) \times (TS^5 \oplus \O)=(\O^6) \times (\O^6)
$.
As in \cite{LMW},  
consider the mapping torus $(X^{10} \times S^1)_f=S^5 \times S^5 \times S^1$ associated to 
the identity diffeomorphism  (id, id) on the two factors in $X^{10}$, 
and let ${{\Delta}}: S^5 \hookrightarrow S^5 \times S^5$ be the diagonal map
 $\Delta (x)=(x,x)$. The normal bundle of $\Delta(S^5) \times \{{\rm pt}\}$ 
 in $S^5 \times S^5 \times S^1$ has a natural trivialization $TS^5\oplus \O=\O^6$.
 The triviality of the normal bundle allows us to use surgery to attach the handle made up of the 
product of two 6-disks
 $\mathbb{D}^6\times \mathbb{D}^6$ to $S^5\times S^5 \times S^1 \times [0,1]$ by 
 gluing $S^5\times \mathbb{D}^6$ to a neighborhood of 
 $\Delta(S^5) \times \{{\rm pt}\} \times \{1\}$ in $S^5 \times S^5 \times S^1 \times \{1\}$ 
 via the trivialization. In the resulting manifold $W$, we have embedded the 6-disk 
 $\mathbb{D}^6$ with 
 trivial normal bundle and boundary the diagonal 5-sphere $\Delta(S^5)$. 
Let $p$ be a base point of $S^5$. Then the quadratic form corresponding to the diagonal
map is 
\bea
q_X(\Delta_*(S^5))&=&q_X\left( [S^5\times p] + [p\times S^5] \right)
\nonumber\\
&=&q_X\left( [S^5 \times p] \right) + q_X([p\times S^5]) + \phi_X \left([S^5 \times p], [p\times S^5] \right)\;.
\eea 
The two quadratic forms on the right hand side are equal as we can exchange the two factors
by a Spin preserving diffeomorphism. This implies that the left hand side is equal to 
the intersection form which is odd, that is $q_X(\Delta_*(S^5))=1$ (mod 2).

%%%%%%%%%%%%%%
 \subsection{Diffeomorphisms preserving the quadratic structure}
 %%%%%%%%%%%
 \label{sec diffq}
 We would like to (also) preserve the quadratic form, as we mentioned above.
Ultimately, what we need  is to quotient the space of Riemannian metrics 
by an intersection of diffeomorphisms preserving the Spin structure
(or Spin diffeomorphisms) with diffeomorphisms  
 preserving quadratic refinement.
  One way to ensure we get the latter  
 is to have the diffeomorphism induce an isometry on the quadratic forms.

 \paragraph{Isometric quadratic forms.}
Let $q_1$ and $q_2$ be two quadratic forms. An {\it isometry} 
$f: q_1 \to q_2$ is a linear map between the underlying vector spaces
$V_{q_1} \to V_{q_2}$ such that $q_1(x)=q_2(f(x))$ for all $x \in V_{q_1}$. 
If such an isometry exists, we write $q_1 \simeq q_2$ and say $q_1$ and $q_2$ 
are isometric. 

\paragraph{Preserving the quadratic refinement.}
We have seen in section \ref{sec 10} that the $\Z$-intersection
pairing leads to a corresponding $\Z_2$-intersection pairing, which 
can be identified with a map $q: H_5(X^{10};\Z) \to \Z_2$
satisfying relation \eqref{eq q mod 2}.
If $f: X^{10}\to X^{10}$ is a Spin preserving diffeomorphism 
of $(X^{10}, \omega)$, then by naturality of the construction in Ref. \cite{LMW}
(of which we will make more use in section \ref{sec mid}),
we have $q(f_*(x))=q(x)$ for all $x$ in $H_5(X^{10};\Z)$. 
Therefore, the diffeomorphism $f$ preserves the 
quadratic refinement.

\paragraph{Preserving quadratic forms.}
There are various invariants that are defined to determine whether 
quadratic forms over an arbitrary field $\F$ are isometric. These invariants
live in Galois cohomology $H^i\F$ corresponding to the field $\F$. The following 
invariants correspond to cohomology classes of ascending degrees, starting 
from degree 0. They are all defined on the Witt group $W\F$ of the field $\F$. 
In addition, they behave like obstructions in the sense that
the $j$-th invariant is a homomorphism when restricted to the kernel of the
$(j-1)$-th invariant. There invariants ${\rm Inv}_j(q)$ are:

\begin{enumerate}
\item {\it Dimension:} In order to get an invariant that vanishes on hyperbolic forms, one
considers 
\(
{\rm Inv}_0(q)=\dim q ~({\rm mod}~ 2) \in \Z_2=H^0\F\;.
\)

\item {\it Discriminant:} For $q$ a quadratic form of dimension $n$, 
\(
{\rm Inv}_1(q)=(-1)^{n(n-1)/2}{\rm det} q \in \F^\times/\F^{\times 2}=H^1\F\;.
\)
\item {\it Clifford invariant:} This is an invariant 
of the Clifford algebra or the even Clifford
algebra, depending on the dimension invariant, and takes values in the 
2-exponent part of the Brauer group of $\F$
\(
{\rm Inv}_2(q)=
\left\{
\begin{array}{ll}
\left[C\ell(q)\right] \in {}_2{\rm Br}(\F) & {\rm if~} \dim q {\rm ~ is~even};\\
\left[C\ell_0(q)\right] \in {}_2{\rm Br}(\F) & {\rm if~} \dim q {\rm ~ is~odd}.
\end{array}
\right.
\)
\end{enumerate}
In general there are more invariants, ${\rm Inv}_n: \ker {\rm Inv}_{n-1} \to H^n\F$
for all $n\geq 0$;
however not all are needed due to a truncation process. Then the problem of 
deciding whether two quadratic forms $q_1, q_2$ over $\F$ are isometric 
can be solved by computing cohomology classes. First, one checks that 
$\dim q_1=\dim q_2$. If this holds then one checks that $q_1-q_2$ is hyperbolic.
This process can be tested by successively ensuring that ${\rm Inv}_i(q_1 -q_2)=0$
for $i$ running over the ordered set $\{0,1, \cdots d\}$, where the process 
truncates at $i=d$ for $2^d \leq \dim q_1 + \dim q_2$ via the 
so-called Arason-Pfister Hauptsatz (see \cite{Tig}). Note that for $\F=\Z_2$, the 
dimension and the Arf invariant form a complete invariant. 

\vspace{3mm}
 We have seen in relation \eqref{eq const} how (essentially) the 
sum of the first two invariants above-- but for the bilinear form-- 
is constrained by the value of the bilinear form at a characteristic element. 

\vspace{3mm}
We will need the following related concept. 

\paragraph{
%%First invariant: 
Isometric structure.}
An isometric structure over $R=\Z$ or $\Z_2$ is a triple 
$(V, s, {\mathcal{I}})$, where 

\noindent $\bullet$ $V$ is a free finite-dimensional $R$-module.

\noindent  $\bullet$ $s: V \times V\to R$ an antisymmetric unimodular bilinear form.

\noindent $\bullet$ ${\mathcal{I}}: V \to V$ is an isometry of $(V,s)$ into itself, i.e. for all $x,y$ in $V$, 
$s(x,y)=s({\mathcal{I}}(x), {\mathcal{I}}(y))$. 
 
\noindent For us $V$ is the middle cohomology, $s$ is the intersection pairing,
and $h$ is the isometry of the intersection pairing (later this will be induced
from a diffeomorphism $f$ as $f_*$ on the homology). 
The sum of two isometric structures is defined by the orthogonal direct
sum 
$(V_1, s_1, {\mathcal{I}}_1) + (V_2, s_2, {\mathcal{I}}_2)=
(V_1 \oplus V_2, s_1 \oplus s_2, {\mathcal{I}}_1 \oplus {\mathcal{I}}_2)$.
The abelian group of equivalence classes $[V, s, {\mathcal{I}}]$ of isometric structures 
denoted by 
\footnote{The -1 subscript refers to antisymmetric.}
$W_{-1}(\Z;R)$, the {\it Witt group} of antisymmetric 
structures over $R$.  For $R$ equal to $\Z$ or $\Q$, the Witt group is 
infinite-dimensional and is given by 
$
W_{-1}(\Z; \Z)\cong \Z^\infty \oplus \Z_2^\infty \oplus \Z_4^\infty$.
The torsion-free part is detected by the equivariant signature
and the torsion is related to number theoretic invariants
that we will not consider here (see \cite{Nu}).

\vspace{3mm}
Now for $(X^{10}, f)$ a diffeomorphism of a 10-dimensional closed manifold, the 
intersection form $s$ on $H_5(X^{10};\Z)/{\rm Tor}$ is antisymmetric and unimodular 
by Poincar\'e duality. The diffeomorphism $f$ induces an isometry 
$f_*: H_5(X^{10};\Z)/{\rm Tor} \to H_5(X^{10};\Z)/{\rm Tor}$. The isometric 
structure $I(X^{10}, f)$ of a diffeomorphism $(X^{10}, f)$ is defined as 
\(
[H_5(X^{10};\Z)/{\rm Tor}, \phi_X^*, f_*] \in W_{-1}(\Z;\Z)\;.
\)
%Here $s$ is the intersection form on $X^{10}$. 
This equivalence class in the Witt group is a cobordism invariant. 
See \cite{Kr} for more details.
%\vspace{3mm}
%The equivariant signature of $(X^{10}, f)$ is defined as 
%${\rm sign}(X^{10},f)={\rm sign}(X^{10},f)$. 
%If $f$ is contained in a compact Lie group acting on $X^{10}$ then this 
%coincides with the equivariant signature of Atiyah-Singer \cite{ASIII}. 
%For $f_*={\rm id}$ we get the Hirzebruch signature of $X^{10}$
%(which is zero for dimension reasons). 
%

\paragraph{Preserving Spin structure vs. preserving quadratic structure.}
As mentioned at the beginning of Section \ref{sec dif} and the  
introduction to the current Section,
one way to ensure preserving 
both the Spin structure and the quadratic structure is to restrict to those
diffeomorphisms which lie in the intersection of the diffeomorphisms 
preserving the first and those preserving the second.  
There is in fact a map from the set of Spin structures on $X^{4k+2}$ to
the set of quadratic refinements of the mod 2 intersection pairing on 
$H_{2k+1}(X^{4k+2}; \Z_2)$ \cite{Br}. 
The set of Spin structures is $H^1(X; \Z_2)$ and the set of quadratic refinements
is the 2-exponent group ${}_2H^{2k+1}(X^{10}, U(1))$ (cf. \cite{BeM}).
However, this map is 
neither injective nor surjective in general, so that knowing one 
side of the map does not in general tell us about the other in any
complete way. However, in the case of Riemann surfaces, corresponding 
to $k=0$, the map is an isomorphism. What we can do is assume 
that one of the sets is a subset of the other set. For example, we 
can assume an injection $H^1(X^{10}; \Z_2) \hookrightarrow 
{}_2H^5(X^{10}, U(1))$, so that preserving the quadratic refinement
also preserves the Spin structure. Depending on whether the 
number of Spin structures is large, we can also assume an injection
the other way. 
At any rate, as  mentioned in the remarks at the end of section \ref{sec glob},
we do not need to go into such specifications in order to arrive at the 
conclusions on anomaly cancellation.

%%%%%%%%%
\subsection{
%The action of d
Diffeomorphism on (almost) middle 
cohomology in  11 and 12 dimensions}
%%%%%%%%%
\label{sec mid}

We have seen in Section \ref{sec 11} that the 
torsion subgroup in eleven dimensions and the 
corresponding linking pairing, equation \eqref{eq link},
are related to the Arf invariant.
In this section we will 
see how both data, the torsion subgroup $T_5((X^{10}\times S^1)_f)$
and the linking pairing $L$, can be described in terms of the 
induced mapping $f_*: H_5(X^{10};\Z) \to H_5(X^{10};\Z)$ on the 
middle-dimensional cohomology of the base 10-manifold. 
We will make use of the construction in Ref. \cite{LMW}.

%\vspace{3mm}
\paragraph{Extension to the mapping torus.}
Consider an element $y \in H_5(X^{10};\Z)$. We would like to 
see how much $y$ changes under the action of $f_*$, that is to 
the new element $f_*y$. To that end, consider the difference $y-f_*y$,
represented by the action of the map $(1-f_*)$ on the element
$y$. Requiring this difference to be 
zero might be too much to ask as then we would be saying that these element are
actually invariant. However, we would like to do something close, namely 
consider the above difference to be a nonzero 
multiple of a nontrivial element $x$ in $H_5(X^{10};\Z)$. Hence we 
consider a summand in $H_5(X^{10};\Z)$ given by 
\(
\cA=\{
x \in H_5(X^{10};\Z)~|~Nx=y-f_*y {\rm ~for~some~nonzero~integer~}
N {\rm ~ and~some~} y \in H_5(X^{10};\Z) 
\}\;.
\label{def A}
\)
On this group, define the rational bilinear pairing $\cB: \cA \times \cA \to \Q$ by the formula
\(
\cB(x_1, x_2)=\frac{1}{N} \cdot \phi_X^*(x_1, x_2)\;,
\)
where $Nx_2=y_2-f_*(y_2)$ and $\phi_X^*$ is the intersection pairing on $X^{10}$.  
The image of the map $(1-f_*): H_5(X^{10};Z) \to H_5(X^{10};\Z)$ 
is contained in $\cA$ and the quotient $[\cA/{\rm im}(1-f_*)]$ is a finite 
abelian group, i.e. a lattice. In fact, the inclusion $\iota: X^{10}
\hookrightarrow (X^{10}\times S^1)_f$ of $X^{10}$ into $X^{10}\times 0$
 leads to an isomorphism of torsion groups $[\cA/{\rm im}(1-f_*)]\cong T_5((X^{10}\times S^1)_f)$.
 Indeed, consider homology long exact sequence
 \(
 \xymatrix{
 \ar[r]
 &
 H_5(X^{10}; \Z)
 \ar[r]^{(1-f_*)}
 &
 H_5(X^{10}; \Z)
 \ar[r]^{i_*}
 &
 H_5(Y^{11}; \Z)
 \ar[r]
 &
 H_4(X^{10}; \Z)
 \ar[r]^{(1-f_*)}
 &
 H_4(X^{10}; \Z)
 %\ar[r]
 }\;.
 \)
 Since $H_5(X^{10}; \Z)$ is assumed to be torsion-free, then so is $H_4(X^{10}; \Z)$ by
 Poincar\'e duality. 
 \footnote{See Section \ref{sec coh ho}.}
 This implies that the torsion subgroup $T_5(Y^{11})\subset H_5(Y^{11}; \Z)$ 
 does not get any contribution from elements in $\ker (1-f_*): H_4(X^{10}; \Z) \to H_4(X^{10}; \Z)$.
 This gives the desired result.

 \vspace{3mm}
 Now the bilinear form $\cB$ on the set $\cA$ induces a corresponding 
 bilinear form $\cB'$ on this torsion group 
\(
\cB': [\cA/{\rm im}(1-f_*)] \times [\cA/{\rm im}(1-f_*)] \to \Q/\Z\;,
\label{eq b'}
\)
which is exactly the linking pairing $L$ on $T_5(Y^{11})$; cf. equation 
\eqref{eq link}.

\vspace{3mm}
Consider a quadratic refinement $q$ which is compatible with $f$ in the sense
that $q(f_*(x))=q(x)$ for all $x$ in $H_5(X^{10};\Z)$ (cf. Section \ref{sec diffq}). 
Associated to this quadratic refinement there is a mapping
$
\tilde{Q}[q]: \cA \to \Q/\Z$,
defined by $\tilde{Q}[q] (x)=\frac{1}{2}\cB (x,x) + j(q(x))$, where $j: \Z_2 \hookrightarrow
\Q/\Z$ represents the inclusion $j(1)=\frac{1}{2}$, $j(0)=0$. 
From the general construction of Ref. \cite{LMW}, 
the mapping $\tilde{Q}$ induces a quadratic refinement
\(
Q[q]: [\cA/{\rm im}(1-f_*)] \to \Q/\Z
\label{eq Qq}
\)
of the nonsingular pairing $\cB'$ in \eqref{eq b'} (which coincides with the 
linking pairing $L$  on the mapping torus given in \eqref{eq link}). 
Recall that this quadratic refinement was defined 
solely from the  the map $f_*$ and the quadratic form $q$, both on 
the basic middle homology $H_5(X^{10};\Z)$.

\paragraph{Extension to twelve dimensions.}
Consider $Y^{11}=(X^{10}\times S^1)_f$ as the boundary of a compact smooth oriented
12-manifold $Z^{12}$. We need to choose  a {\it relative Wu class}
$v'\in H^6(Z^{12}, Y^{11};\Z_2)$ whose restriction $v'|_{Z^{12}}$ is the 
6th Wu class $v_6(Z^{12})\in H^6(Z^{12};\Z_2)$, and which is compatible with 
the quadratic refinement $Q$. The most convenient choice is  $v'=0$.
\footnote{See the end of Section \ref{sec char v} as well as Section \ref{sec Wu} 
for a discussion on Wu classes.} 
But then for all relative homology classes $b \in H_6(Z^{12},Y^{11};\Z)$ with 
$\partial b$ a torsion class in $H_5(Y^{11};\Z)$ we should have
\(
Q[q](\partial b)=-\frac{1}{2} \phi_Y(b, \tilde{b}) ~({\rm mod}~\Z)~~{\rm in~} \Q/\Z\;,
\label{eq comp}
\)
where $\tilde{b}$ is some choice of rational class in $H_6(Z^{12};\Q)$ which has
the same image as $b$ in the relative rational homology group $H_6(Z^{12}, Y^{11};\Q)$.
This can be checked explicitly using chains \cite{LMW}.

%%%%%%%
\subsection{Description via the Rochlin invariant}
%%%%%%
\label{sec Roch}

In this section we will see how the expression for the anomaly in type IIB 
involving the combination of eta invariants on the mapping torus 
(cf. expressions  \eqref{eq hol 3} and \eqref{eq eta tot})
is encoded in the 
Rochlin invariant. 
This will be an overview and an application of the mathematical 
results in  \cite{ML} \cite{LMW} \cite{FK}.

\vspace{3mm}
On the Spin manifold $Y^{11}=(X^{10}\times S^1)_f$ there exists a well-defined 
$\Z_{16}$-invariant $R(Y^{11})$ given by the formula \cite{ML}
\(
R(Y^{11})=\sigma (Z^{12}) \mod 16 ~ \in \Z_{16}\;.
\)
This Rochlin invariant is well-defined; this follows from the Novikov additivity of the
signature and the divisibility in the closed case, i.e. Ochanine's result \cite{Och}
that  the signature $\sigma (Z^{12})$ on the intersection pairing 
$\phi^*_Z : H_6(Z^{12};\Z) \otimes H_6(Z^{12};\Z) \to \Z$ 
of the middle-dimensional homology of
a compact {\it closed} Spin manifold 12-manifold $Z^{12}$
is divisible by 16. 
\footnote{So had $Z^{12}$ been closed then showing absence of 
global anomaly would have been straightforward.}
The value of the Rochlin invariant modulo 8 is independent of the 
choice of Spin diffeomorphism $F=(f,b)$ covering $f$ and only 
depends on data related to the middle (co)homology, namely:
\begin{enumerate}
\item The quadratic mapping $q_\omega: H_5(X^{10};\Z)\to\Z_2$ defined by the 
Spin structure $\omega$. This is a 
quadratic refinement of the intersection pairing on $X^{10}$ constructed by 
Brown \cite{Br}. 
\item The induced map $f_*: H_5(X^{10};\Z) \to H_5(X^{10};\Z)$  on the 
middle-dimensional homology of $X^{10}$. 
\end{enumerate}
Since $Y^{11}$ with its Spin structure $\omega$ always bounds, then 
from \cite{LMW}, the Rochlin invariant is given by 
\(
R(Y^{11}, \omega)= \frac{1}{2}\eta(Y^{11}, \cS) 
+ 4[ h(Y^{11};D_{TY}) + \eta(Y^{11};D_{TY})]
-16 \eta (Y^{11};D) \mod~16\;.
\label{def roch}
\)
%\attn{Check numbers !!!}

\vspace{3mm}
We now consider the Rochlin invariant in the presence of some structure. 
From \cite{BM}, we have for the relative cohomology
$H^6(BSO, BSO[v_6];\Z_2)\cong \Z_2$ and so this group
contains a unique nonzero element $v$. Let $g: Z^{12}\to BSO$ denote the classifying map of
the stable normal bundle of $Z^{12}$. The pair of mappings 
$(g, \tilde{\nu}): (Z^{12}, X^{10}) \to (BSO, BSO[v_6])$
can be used to pullback $v$ to a cohomology class $\tilde{v}$
in the relative cohomology group $H^6(Z^{12}, X^{10};\Z_2)$;
indeed, we know that $X^{10}$ admits a Wu-structure
(see end of Section \ref{sec mid}).

\vspace{3mm}
From the point of view of the mapping torus $Y^{11}=(X^{10}\times S^1)_f$, we need
to consider the corresponding pairing $L$ on the torsion subgroup $T^5(Y^{11})$
as well as the quadratic refinement $Q_L$ (see expression \eqref{eq QL}). 
Using the quadratic refinement $Q_L$, one can assign to such a class $\tilde{v}$ 
a modulo 8 invariant $\tilde{v}_Q^2$ such that the following relation holds \cite{ML}
\(
\tilde{v}_Q^2 - A(Y^{11}, Q_L)=\sigma (Z^{12}) ~~\mod 8\;.
\)
If $Z^{12}$ is taken to be a Spin manifold then the maps to $BSO$ and 
$BSO[{v_6}]$ factor through $B{\rm Spin}$, so that the pair 
$(g, \tilde{\nu})$ induces a trivial map between relative 
cohomology groups (see diagram \eqref{diag BSO}) and in this case we
have $\tilde{v}=0$ and $\tilde{v}_Q=0$.  

\vspace{3mm}
Let us consider $X^{10}$ to be Spin with a Spin structure $\omega$.  
From Ref. \cite{Br}, the Spin structure $\omega$ gives a canonical refinement 
$q$ of the $\Z_2$-intersection pairing, that is
\bea
q: H_5(X^{10}; \Z) &\longrightarrow& \Z_2
\nonumber\\
q(x+y) - q(x) - q(y)&=& \phi^*_X (x, y)~~({\rm mod}~2)\;.
\eea
Then in this case where all manifolds are Spin,
and using \cite{LMW}, we have that the 
 Rochlin invariant, the Arf invariant, and the signature are related as 
\(
R(Y^{11})=\sigma(Z^{12})=-A(Y^{11}, Q_L) ~~({\rm mod}~8)\;.
\)

\paragraph{The Rochlin invariant in terms of the Arf invariant.}
We have seen towards the end of Section \ref{sec mid} that 
a quadratic refinement can be constructed on the mapping torus
starting from the action of the diffeomorphism on the middle cohomology
of the base 10-manifold $X^{10}$, via $f_*$, and from the corresponding 
quadratic form $q$. This quadratic refinement satisfies some 
compatibility conditions spelled there (cf. equation \eqref{eq comp}).
Thus with compatibility, 
via \cite{BM}, the Rochlin invariant of the mapping torus is given by 
the Arf invariant of this quadratic form $Q[q]$ in \eqref{eq Qq}, that is
\(
R((X^{10}\times S^1)_f, \omega')=-A(Q[q])~~({\rm mod}~8)\;.
\)

%\vspace{1cm}
%If $(Y^{11},w)$ is a Spin boundary, so that the Rochlin invariant 
%$R(Y^{11}, w)$ is defined, then \cite{LMW}
%\(
%R(Y^{11}, w)=4 {\rm Arf}(Y^{11}, w) ~\in ~\Z_8\;.
%\)

\paragraph{Example 9: Products with $b_5=0$.}
We can consider the case of product manifolds with the possibility that 
diffeomorphisms on one or more of the factors are trivial. 
%\paragraph{Case $H_5=0$.} 
Take $X^{10}$ to be the product manifold $T^2 \times \H P^2$
of a two-torus with the quaternionic projective plane.
Take $\alpha:=(T^2, \omega, f, h) -(T^2, \omega, {\rm id}, h')$ where 
$f$ is given by $\binom{1~1}{0~1}$ and $\omega$ is the standard Spin 
structure of $T^2=\R^2/\Z^2$ (with nontrivial Arf invariant),
$h$ an appropriate homotopy and $h'$ a constant homotopy. Then the
Rochlin invariant is even,
$R(\alpha)\equiv 0$ (mod 2). 
Now 
consider the quaternionic projective plane $\H P^2$ and 
take $\beta:=(\H P^2, \omega, {\rm id}, h)$ with any Spin 
structure, identity diffeomorphism and constant homotopy. 
Then the product 
$(T^2 \times \H P^2)_{f\times {\rm id}}=(T^2 \times S^1)_f \times \H P^2$ is a 
Spin boundary of $M^4 \times \H P^2$ if $\partial M^4=(T^2 \times S^1)_f$. 
Since the signature is multiplicative, the Rochlin invariant of the 
product is \cite{FK}
$R(\alpha \times \beta)={\rm sign}(M^4 \times \H P^2)
={\rm sign}(M^4)\cdot {\rm sign}(\H P^2)$,
which is equal to $R(\alpha)$ since the signature of $\H P^2$ is 1. 
We will consider this example further in section \ref{sec Kreck}.

\vspace{3mm}
Note that this allows us to make use of the transparent 
Riemann surface case, for which there is a one-to-one 
correspondence between the set of Spin structures and the
set of quadratic refinements. The general case is reduced
to this particular case by taking $X^{10}=\Sigma_g \times \R^8$,
where $\Sigma_g$ is a Riemann surface of genus $g$.

%%%%%%%%%%%%%%%%
%\subsubsection{Change of Spin structure}
%%%%%%%%%%%%%%

\paragraph{Variation of Spin structure on $X^{10}$ and the Arf invariant.}
Now we consider the situation where $X^{10}$ has (at least) two Spin structures.
This means that $X^{10}$ has to satisfy $|H^1(X^{10};\Z_2)| \geq 2$. 
Let $f: X^{10} \to X^{10}$ be an orientation preserving diffeomorphism 
which preserves two Spin structures $\omega_1, \omega_2$ on $X^{10}$. 
 Lift $f$ to two Spin diffeomorphisms $F_1=(f, b_1)$ and $F_2=(f, b_2)$ which 
preserve the Spin structures $\omega_1$ and $\omega_2$, respectively. 
Let $\omega'_1$, $\omega'_2$ be the Spin structures on $(X^{10} \times S^1)_f$
corresponding to these two choices $F_1$ and $F_2$. 
In this situation, the Rochlin invariant of the difference 
$R[(Y^{11}, \omega'_1) - (Y^{11}, \omega'_2)]$ is always defined since
$Y^{11}$ is always the Spin boundary of some 12-manifold $Z^{12}$.

Now let $q_1, q_2: H_5(X^{10};\Z) \to \Z_2$ be the quadratic refinements of the 
intersection pairing determined by the Spin structures $\omega'_1$, $\omega'_2$, respectively. 
We need to look at the value of the difference $q_2(y) - q_1(y)$  inside $\Q/\Z$ via the 
embedding $j: \Z_2 \hookrightarrow \Q/\Z$, 
for any element $y$ in the group $\mathcal{A}$, defined in \eqref{def A}. 
In fact, there is a unique element 
$z$ in the quotient $[\mathcal{A} /{\rm im}(1-f_*)]$ such that $j[q_2(y)-q_1(y)]$ 
coincides with the bilinear form $\cB'(y,z)$, defined in \eqref{eq b'},
 for all $y$ in $\mathcal{A}$. Note that $\cB'$ is 
nonsingular, which is compatible with $j(0)=0$ and the fact that we take 
$q_1$ and $q_2$ to be distinct. Then, building on \cite{LMW},
 the Rochlin invariant of the difference is essentially given by the difference of the Arf invariants
 of the corresponding quadratic forms
\(
R[(Y^{11}, \omega'_1) - (Y^{11}, \omega'_2)]=A[Q(q_1)] - A[Q(q_2)]~~({\rm mod}~8)\;.
\)  
%It also holds that
%
%\vspace{3mm}
%$4\left(A[Q(q_1)] - A[Q(q_2)]\right)=0 \in \Q/\Z$.
%
%
%\attn{So in simply connected case we have only one Spin 
%structure and so we would
%then only require that the Arf invariant is divisible by 8}

\paragraph{Variation of Spin structure and the Ochanine invariant.}
Let $(X^{10},f)$ be a fixed connected Spin manifold and $f$ a Spin diffeomorphism. 
Then there are exactly two homotopy classes of homotopies 
from $f\circ \omega$ to $\omega$. In particular, there are exactly two Spin 
structures on the mapping torus corresponding to the identity 
diffeomorphism  $(X^{10} \times S^1)_{\rm id}=X^{10} \times S^1$, 
and let $\overline{h}$ be the one
nontrivial on $S^1$. Then $X^{10} \times S^1$ is also a Spin boundary with respect to 
$\overline{h}$. 
%\paragraph{The Ochanine invariant.} 
The Ochanine invariant \cite{Och} is defined in our setting as 
\(
O(X^{10}, \omega):= R(X^{10}, \omega, {\rm id}, \overline{h}) \in \Z_{16}\;.
\)
Note that $O(X^{10}, \omega) \in 8\cdot \Z_{16}\cong \Z_2$ since 
$2O(X^{10}, w)=R(2(X^{10}, \omega, {\rm id}, \overline{h}))=0$.
This invariant is always divisible by 8.
Next, let $h$ and $h'$ be representatives of the two homotopy classes of homotopies 
joining $\omega$ to $\omega \circ f$. Then, using \cite{FK}, we have that the variation 
of the Rochlin invariant is given by the Ochanine invariant of the base
\(
R(X^{10}, \omega, f, h)- 
R(X^{10}, \omega, f, h')= O(X^{10}, \omega)\;.
\label{eq roch och}
\)
In fact, as can be deduced from Ref. \cite{Da},
both of the above variations are zero mod 8. 
Applications of this invariant to the partition function in M-theory is given in Ref. 
\cite{S2}.   

\paragraph{Effect of torsion in middle (co)homology.}
To which extent is the Rochlin invariant $R$ 
determined by the induced map on $H_5(X^{10};\Z)$?
This will depend on whether or not torsion is present.
The formulation in  \cite{LMW} gives a formula for 
$R$ (mod 8) in terms of the induced map, 
if $H_5(X^{10};\Z)$ is torsion-free. As argued in \cite{FK} 
(in more generality than what we need) this 
condition cannot be dropped so that there cannot be a formula
depending only on the induced map $f_*$ on $H_5(X^{10}; \Z)$ and the 
Spin structure. An example which illustrates this is given towards the
end of Section  \ref{sec Kreck}.
%Let $T_5(X^{10})$ denote the torsion subgroup of the 5th 
%homology group of $X^{10}$. 
%From \cite{BM} 
There is a relative cohomology class $x \in H^6(Z^{12}, Y^{11};\Z_2)$ such that
\(
\langle x \cup x, [Z^{12}, Y^{11}] \rangle
+
\dim_{\Z_2} (T^6(Y^{11})\otimes \Z_2)
\equiv 
\sigma (Z^{12}) \mod 2\;. 
\)
%But, as we saw in Section \ref{sec Wu}, 
%the Wu class of a Spin 10-manifold vanishes. 
By Poincar\'e  duality, $T^6(Y^{11})\cong T_5(Y^{11})$.
Furthermore, if we set $x=0$, then  
the Rochlin invariant in this case can be calculated only mod 2 
as 
\(
R(X^{10}, w)\equiv \dim_{\Z_2}(T_5(X^{10})\otimes \Z_2) \mod 2.
\)

%\attn{So this does not seem to have an effect on the anomaly 
%since 2.16=8}
%
%\attn{or maybe it is about even vs odd torsion?}
%
%\vspace{2mm}
%\noindent {\bf 2.} {\it The torsion-free case.}

%%%%%%%%%%%%%%%%%%%%%%
\subsection{Description via cobordism invariants related to the Rochlin invariant} 
%{Cobordism of diffeomorphisms and the eta invariants mod 16}
%%%%%%%%%%%%%%
\label{sec Kreck}

We have seen that the global anomaly formula involves the division 
of this linear combination of eta invariants by 8. It is then natural to ask
whether this division leads to an integer or just a rational number. 
%It will turn out that this ratio is not only an integer, but in fact an even integer. 
This makes a direct use of the results in \cite{FK}
as well as the  constructions in \cite{Kr}.
One of the byproducts is an explanation of the extension from the 
circle to the Riemann surface in diagram \eqref{diag 1} and the 
discussion around it.

\paragraph{Cobordism of diffeomorphisms.} 
The cobordism group of $m$-dimensional 
diffeomorphisms $\Delta_m$
is the cobordism group 
of differentiable fiber bundles over $S^1$ with 
$(m+1)$-dimensional total space and is given by the 
mapping torus. 
In the case of $X^{10}$ in type IIB, we have to consider 
$\Delta_{10}$, which is not finitely generated nor  
finite-dimensional  (even rationally). It is natural to ask
how the cobordism group of 10-dimensional diffeomorphisms 
$\Delta_{10}$ is related to other `more common' cobordism groups. 
To answer this question, we would like to describe three homomorphisms from $\Delta_{10}$.
%The first two are simple to describe:

\vspace{2mm}
\noindent {\bf 1.} There is an obvious homomorphism from the cobordism group
of diffeomorphisms $\Delta_{10}$ to the cobordism group $\Omega_{10}\cong \Z_2$
of closed oriented 10-manifolds  
given by forgetting the diffeomorphism and considering only the cobordism 
class of the underlying 10-manifold, that is $[X^{10},f] \mapsto [X^{10}]$. 

\vspace{2mm}
\noindent {\bf 2.} The mapping torus construction raises the dimension by one, and 
there is a homomorphism from $\Delta_{10}$ to $\Omega_{11}\cong \Z_2$, the 
cobordism group of closed oriented 11-manifolds, given by 
$[X^{10}, f]\mapsto [(X^{10}\times S^1)_f]$. The image of this map, denoted 
$\widehat{\Omega}_{11}$, 
coincides with the kernel of the Hirzebruch signature operator $\tau$ because the 
total space of a fibration over the circle has a vanishing signature \cite{Nu-fib}.

\vspace{2mm}
%We now consider the the third homomorphism.
\noindent {\bf 3.} The isometric structure (of section \ref{sec diffq})
leads to the surjective homomorphism 
$I(X^{10},f): \Delta_{10} \to W_{-1}(\Z;\Z)\cong \Z^\infty \oplus \Z_2^\infty \oplus \Z_4^\infty$.
This  third homeomorphism is much more involved than the first two and requires the use of the 
 the Neumann invariant (see below).

\vspace{3mm}
Putting the three homomorphisms together, the `total' homomorphism
$
\Delta_{10}\to W_{-}(\Z;\Z) \oplus \Omega_{10} \oplus \Omega_{11}$,
mapping
$\left[X^{10},f\right]$  to $\left( I(X^{10},f), [X^{10}], [(X^{10}\times S^1)_f]\right)$,
is an isomorphism \cite{Kr}. Therefore, the cobordism group of diffeomorphisms 
of oriented 10-manifolds $X^{10}$ is 
$\Delta_{10}\cong \Z^\infty \oplus \Z_2^\infty \oplus \Z_4^\infty \oplus \Z_2 \oplus \Z_2$.

\paragraph{The case of $X^{10}$ with extra structure.}
The above discussion was for general manifolds with no special structure,
i.e. for $X^{10}$ only oriented. 
We will mainly be interested in the Spin case, that is adding a Spin structure
to the above discussion. However, we could also add other relevant
structures, either more refined such as a String structure, or more
crude such as a framing.  
Generally, if $\mathbb{B}$ is e.g. BSpin, BString, 
BU,  B1,
corresponding to Spin structure, String structure, almost complex
structure, and framing, respectively,
 then \cite{Kr} the kernel of the 
homomorphism 
\(
\Delta_{10}^\mathbb{B} \longrightarrow W_{-}(\Z;\Z) \oplus \Omega_{10}^\mathbb{B} 
\oplus \Omega_{11}^{\mathbb{B}}
\)
is a subgroup of $\Z/\tau(\mathbb{B},12)\Z$, where the denominator is the smallest positive 
signature of a closed 12-dimensional $\mathbb{B}$-manifold. We can consider the following cases:
%\attn{diffeomorphisms preserving quadratic forms}

\vspace{2mm}
\noindent {\bf 1.}  $X^{10}$ {\it is  Spin}: If the 10-manifold $X^{10}$ is Spin then we have to consider 
the cobordism group $\Delta_{10}^{\rm Spin}$ of 10-dimensional 
Spin diffeomorphisms. Here the relevant cobordism groups are in dimension eleven $\Omega_{11}^{\rm Spin}=0$,
and dimension ten $\Omega_{10}^{\rm Spin} \cong \Z_2 \oplus \Z_2 \oplus \Z_2$.

\noindent {\bf 2.} {\it $X^{10}$ is String}: Here the relevant cobordism groups are 
$\Omega_{10}^{\rm String}=\Z_2 \oplus \Z_3$ and $\Omega_{11}^{\rm String}=0$. 

\noindent {\bf 3.} {\it $X^{10}$ is framed}:  Many examples that we considered, 
including $S^5 \times S^5$ are
in fact framed manifolds. 

%However what matters for the Rochlin invariant is a summand...
\vspace{2mm}
\noindent We have, however, concentrated mostly on the Spin case in this paper.

\paragraph{
%Second invariant: 
The Neumann invariant.}
Again, let $Y^{11}=\partial Z$ be the mapping torus 
$(X^{10}\times S^1)_f$ for $X^{10}$. 
Then the Neumann invariant ${\cal N}(X^{10},f)$
is defined to be the signature of the symmetric 
bilinear form on $H_5(X^{10};\Q)$ given by 
\(
(x,y) \longmapsto \phi_X((f_*-f_*^{-1})x, ~ y)\;.
\)
i.e. $\sigma(Z^{12})={\cal N}(X^{10},f)$. 
%is the 
%fiber bundle structure on 
%$Y^{11}$ extends to $Z^{12}$. 
Unlike the isometric structure described in Section \ref{sec diffq} 
above, the Neumann invariant is not 
a cobordism invariant \cite{N3}. 
For example, let $g: S^5 \times S^5 \to S^5 \times S^5$ be the clutching 
function of the sphere bundle of the tangent bundle of the 6-sphere $S^6$. 
Then with respect to the standard basis of $H_5(S^5 \times S^5)$, 
$g_*$ has the matrix description 
$\binom{1~0}{2~1}$ and the intersection form $s$ has the matrix form 
$\binom{~0~1~}{-1~0~}$. This gives the value of the Neumann invariant for the 
mapping torus ${\cal N}(S^5 \times S^5,g)=1$. However, 
the mapping torus $(S^5\times S^5,g)$ is null-bordant since the 
sphere bundle bounds the disk bundle. Therefore, 
$\mathcal{N}$ is not a cobordism invariant.

\paragraph{Example 10.}
Take $X^{10}$ to be the product manifold $T^2 \times \H P^2$. 
Take $\alpha:=(T^2, \omega, f, h) -(T^2, \omega, {\rm id}, h')$ where 
$f$ is given by the matrix $\binom{1~1}{0~1}$ and $\omega$ is the standard Spin 
structure of $T^2=\R^2/\Z^2$ (with nontrivial Arf invariant),
$h$ an appropriate homotopy, and $h'$ a constant homotopy. 
Then the Neumann invariant is odd
 $\mathcal{N} (\alpha) \equiv 1$ (mod 2). 
Then the Neumann invariant for the product reduces to that of the 
2-torus, i.e. $\mathcal{N}(\alpha \times \beta)=\mathcal{N}(\alpha)$. 
We have considered the Rochlin invariant on these manifolds in 
Section \ref{sec Roch}.

\vspace{3mm}
We now make use of an integer invariant that captures the Rochlin invariant 
modulo 16, and hence describes the combination of the eta invariants
appearing in the global anomaly formula.

%%%%%%%%%%%%%%
\paragraph{The Fischer-Kreck  cobordism invariant.} 
Following \cite{FK}, we define the invariant (cf. \eqref{def roch})
\(
R(Y^{11}, \omega)= - \eta(Y^{11}, S) 
+ 4[ h(Y^{11};D_{TY}) + \eta(Y^{11};D_{TY})]
-16 \eta (Y^{11};D) \mod~16
~\equiv S(Y^{11}, \omega)\;.
\)
The cobordism invariant $(Y^{11}, \omega) \mapsto S(Y^{11},\omega) \in \R/\Z$ 
is an integer described as follows (see \cite{FK}).
There is an exact sequence 
\(
0 \to K \to \Delta_{10}^{\rm Spin} \to \Omega_{10}^{\rm Spin}
\oplus \Omega_{11}^{\rm Spin} \oplus W_{-}(\Z;\Z) \to 0
\)
with isomorphism $K \to \Z_{16}$ given by 
$[X^{10}, \omega, f, h] \mapsto R((X^{10}\times S^1) _f, \omega_h) - \overline{\mathcal{N}}(X^{10},f)$ mod 16.
Here $\overline{\mathcal{N}}:=\mathcal{N}$ (mod 16), i.e. is the reduction modulo 16 of the
Neumann invariant described above. 
The sequence in fact splits and, with 
$\Omega_{10}^{\rm Spin}\cong \Z_2 \oplus \Z_2 \oplus \Z_2$ and 
$\Omega_{11}^{\rm Spin}=0$,
 there is an isomorphism 
\(
\Delta_{10}^{\rm Spin} \cong  \Z_2 \oplus \Z_2 \oplus \Z_2 \oplus W_{-}(\Z;\Z) \oplus \Z_{16}\;.
\)
The Witt group $W_{-}(\Z;\Z) $ is described towards the end of Section \ref{sec diffq}.
The new invariant in this case is 
\(
S(X^{10}, \omega , f, h): \Delta_{10}^{\rm Spin} \to \Z_{16}\;.
\)
Note that the exact sequence also shows that every 11-dimensional Spin 
manifold is cobordant to a mapping torus. Therefore,  considering the
Fischer-Kreck invariant for mapping tori in fact covers all Spin 11-manifolds,
since this invariant is a cobordism invariant.

\paragraph{Dependence on homotopy and relation to the Ochanine invariant.}
Let $h$ and $h'$ be representatives of the two homotopy classes of homotopies 
joining the Spin structure $\omega$ to Spin structure $\omega \circ f$. Then, using \cite{FK}, 
we have that the variation 
of the Fischer-Kreck invariant is given by the Ochanine invariant of the base
\(
S(X^{10}, \omega, f, h)- 
S(X^{10}, \omega, f, h')= O(X^{10}, \omega)\;.
\)
This is analogous to the variation of the Rochlin invariant leading to expression \eqref{eq roch och}.
The appearance of the Ochanine invariant in the dual type IIA string theory
has been highlighted in Ref. \cite{S0}.

\paragraph{Example 11: Effect of torsion in middle homology.} 
Consider our standard example, the 10-manifold $X^{10}=S^5 \times S^5$ and let $a$ and $b$  
form a basis of $H_5(X^{10}; \Z)$ which is defined by the embedding of the first
factor and by the diagonal, respectively. The intersection pairing with respect 
to this basis is given by $\binom{~0~1}{-1~0~}$. The mod 2 refinement 
$q$ given by the normal bundle of an embedded sphere is defined by  
$q(x)=0$ is and only if  the normal bundle is trivial. This refinement 
has values $q(a)=0$ and 
$q(b)=1$ on the basis elements. The automorphism of 
$H_5(X^{10}; \Z)$ defined by $a\mapsto a+b$,
$b \mapsto b$ can be realized by a diffeomorphism $f$ of  
$X^{10}$ which keeps a neighborhood $U$ of the diagonal $\Delta (S^5)$ fixed. 
Then in this case the Rochlin invariant is even $R(X^{10}, \omega, f, h)\equiv 0$ (mod 2) 
and the Neumann invariant is $\mathcal{N}(X^{10}, f)=1$.
Now $2b$ can be realized by an embedding $S^5 \hookrightarrow U$
and has a trivial normal bundle, since $q(2b)=0$. This then allows 
us to do surgery using a tubular neighborhood of $2b$ contained in $U$ \cite{FK}.

To that end, consider the resulting 10-manifold
 $\tilde{X}^{10}:=(X^{10}\backslash S^5 \times \mathbb{D}^5)\bigcup \mathbb{D}^6 \times S^4$
and diffeomorphism $\tilde{f}:=f|_{X^{10}\backslash S^5 \times \mathbb{D}^5\cup {\rm id}}$. 
Since the two manifolds $(X^{10},f)$ and $(\tilde{X}^{10}, \tilde{f})$
are Spin cobordant, the difference of their Rochlin invariants and reduced 
Neumann invariants (i.e. essentially the Fischer-Kreck invariants) are equal
$(R-\overline{\mathcal{N}})(X^{10}, f)=(R-\overline{\mathcal{N}})(\tilde{X}^{10}, \tilde{f})$.
Furthermore, we have for the homology groups $H_4(\tilde{X}^{10})
\cong \Z_2 \cong H_5(\tilde{X}^{10})$, so that the middle 
cohomology is torsion. 
Rationally, $H_5(\tilde{X}^{10};\Q)=0$, which implies that that the Neumann invariant
vanishes $\mathcal{N}(\tilde{X}^{10}, f')=0$ so that the Rochlin invariant 
$R(\tilde{X}^{10}, \tilde{f})$ is a generator of $\Z_{16}$. Now for any $n$,
the Rochlin invariant corresponding to an iterated cobordism 
satisfies $R(\tilde{X}^{10}, \tilde{f}^n)=nR(\tilde{X}^{10}, \tilde{f})$.
This means that $\tilde{X}^{10}$ is a Spin manifold with torsion 
middle cohomology such that for any integer $r$ there exists 
a Spin diffeomorphism $(f , h)$ on $(X^{10}, \omega)$ such
that $R(X^{10}, \omega, f, h)\equiv r$ (mod 16).

\vspace{3mm}
The consequences of the above example are two-fold. First that, as stated towards the end of
Section \ref{sec Roch}, that the Rochlin invariant cannot be computed if torsion 
in middle homology is present. Second, aside from torsion, given a value of 
the Rochlin invariant corresponding to a diffeomorphism $f$, one can 
produce any integer multiple of this invariant by considering the 
iterations of $f$.

%\vspace{3mm}
%\paragraph{Closing remarks.}
%%%%%%%%%%%

\vspace{1cm}
\noindent {\large \bf Acknowledgements}

\vspace{2mm}
  The author would like to thank Fei Han for useful discussions on \cite{HL}
  and for kind 
   hospitality at the University of Singapore in January 2011. 
The author would like to thank the American Institute of Mathematics, Palo Alto, 
for hospitality  during the program {\it Algebraic Topology and Physics} in May 2011
as well as IHES, Bures-sur-Yvette, for hospitality in Summer 2011 while this 
project was being completed. 
This research is supported by NSF Grant PHY-1102218.

%%%%%%%%%%%%%%%%%%%%%%%%%%%%%%%%%%%%%%%%%%%%%

\end{document}